\documentclass[%
reprint,
superscriptaddress,
amsmath,amssymb,
aps,
]{revtex4-1}

\usepackage{graphicx}
\usepackage{dcolumn}
\usepackage{bm}
\usepackage[version=4]{mhchem}

\usepackage{color}
\usepackage{siunitx}
\usepackage{hyperref}
\hypersetup{
	final,									
	colorlinks=true,
	breaklinks,	
	pdftitle = {Title},			
	pdfauthor = {Petra Hein},			
	linkcolor = black,							
	urlcolor = blue,						
	menucolor= red,							
	citecolor = blue,						
	filecolor = black,						
	bookmarksopen = true, 		
	bookmarksnumbered = true,	
	pdfpagemode = UseOutlines, 		
	pdfpagelayout = SinglePage,
	pdfmenubar = true,					
	pdftoolbar = true
}

\newcommand{\overbar}[1]{\mkern 1.5mu\overline{\mkern-1.5mu#1\mkern-1.5mu}\mkern 1.5mu}

\begin{document}
	
	\title{Mode-resolved reciprocal space mapping of electron-phonon interaction in the Weyl semimetal candidate \textit{Td}-\ce{WTe2}}

    \author{Petra Hein}
    \email[]{hein@physik.uni-kiel.de}
    \affiliation{Institute of Experimental and Applied Physics, University of Kiel, Leibnizstr. 19, D-24118 Kiel, Germany}
    \author{Stephan Jauernik}
    \affiliation{Institute of Experimental and Applied Physics, University of Kiel, Leibnizstr. 19, D-24118 Kiel, Germany}
    \author{Hermann Erk}
    \affiliation{Institute of Experimental and Applied Physics, University of Kiel, Leibnizstr. 19, D-24118 Kiel, Germany}

    \author{Lexian Yang}
    \affiliation{State Key Laboratory of Low Dimensional Quantum Physics, Department of Physics, Tsinghua University, Beijing 100084, China}
    \affiliation{Collaborative Innovation Center of Quantum Matter, Beijing 100084, China}

    \author{Yanpeng Qi}
    \affiliation{School of Physical Science and Technology, ShanghaiTech University, Shanghai 201210, China}
    \affiliation{Max Planck Institute for Chemical Physics of Solids, N\"othnitzer Str. 40, D-01187 Dresden, Germany}

    \author{Yan Sun}
    \affiliation{Max Planck Institute for Chemical Physics of Solids, N\"othnitzer Str. 40, D-01187 Dresden, Germany}
    
    \author{Claudia Felser}
    \affiliation{Max Planck Institute for Chemical Physics of Solids, N\"othnitzer Str. 40, D-01187 Dresden, Germany}

    \author{Michael Bauer}
    \affiliation{Institute of Experimental and Applied Physics, University of Kiel, Leibnizstr. 19, D-24118 Kiel, Germany}

	\date{\today}
	
	\begin{abstract}
		The selective excitation of coherent phonons provides unique capabilities to control fundamental properties of quantum materials on ultrafast time scales. For instance, in the presence of strong electron-phonon coupling, the electronic band structure can become substantially modulated. Recently, it was predicted that by this means even topologically protected states of matter can be manipulated and, ultimately, be destroyed: For the layered transition metal dichalcogenide \textit{Td}-\ce{WTe2}, pairs of Weyl points are expected to annihilate as an interlayer shear mode drives the crystalline structure towards a centrosymmetric phase. By monitoring the changes in the electronic structure of \textit{Td}-\ce{WTe2} with femtosecond resolution, we provide here direct experimental evidence that the coherent excitation of the shear mode acts on the electronic states near the Weyl points. Band structure data in comparison with our results imply, furthermore, the periodic reduction in the spin splitting of bands near the Fermi energy, a distinct electronic signature of the non-centrosymmetric \textit{Td} ground state of \ce{WTe2}. The comparison with higher-frequency coherent phonon modes finally proves the shear mode-selectivity of the observed changes in the electronic structure. Our real-time observations reveal direct experimental insights into electronic processes that are of vital importance for a coherent phonon-induced topological phase transition in \textit{Td}-\ce{WTe2}.  
	\end{abstract}
	
	\maketitle
	
	\section*{Introduction}
	
\begin{figure}
		\centering
		\includegraphics[width=1\linewidth]{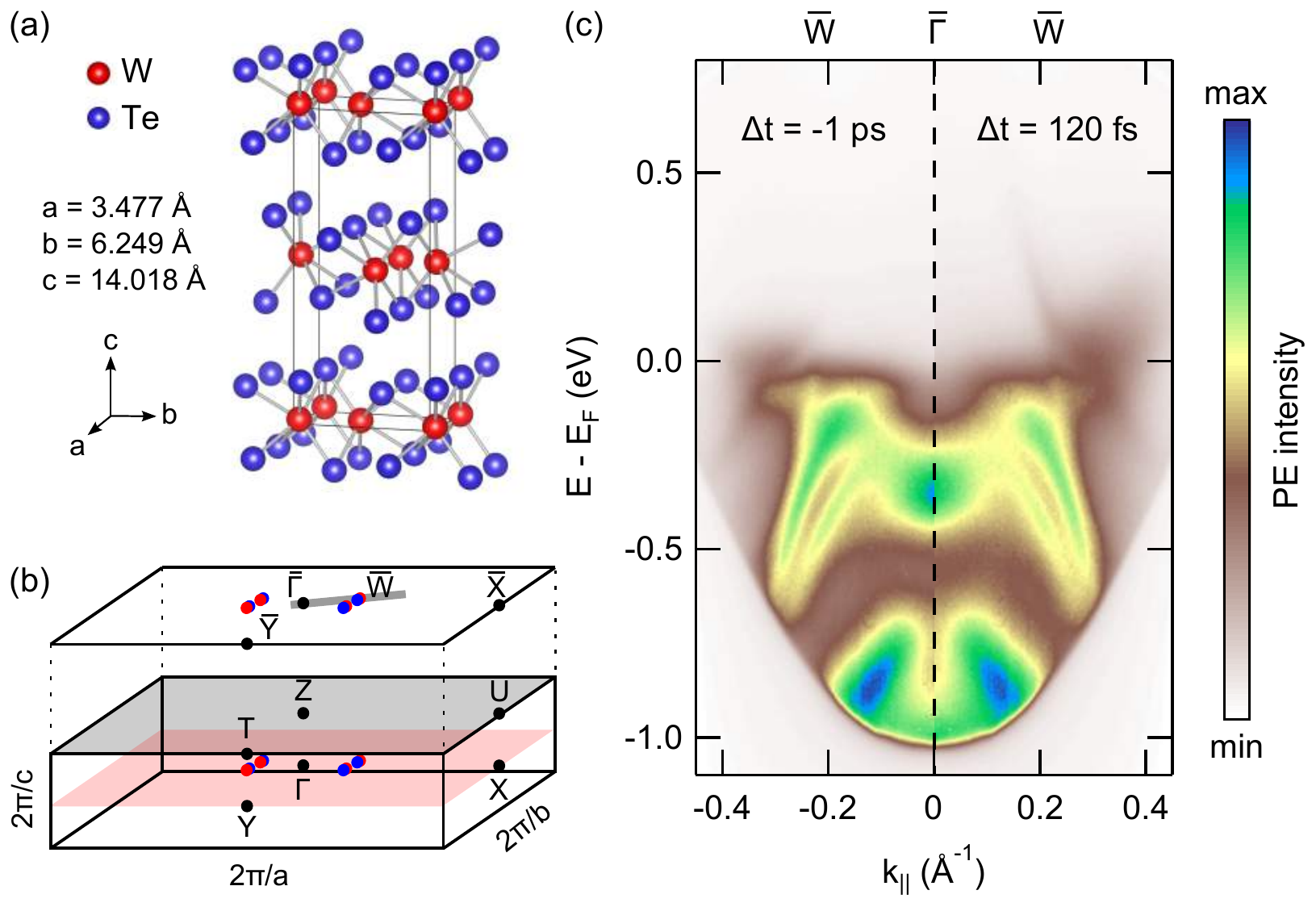}
		\caption{Crystal and electronic structure of \textit{Td}-\ce{WTe2}. (a) Orthorhombic unit cell with experimental lattice constants at $T=\SI{113}{\kelvin}$~\cite{Mar1992}. (b) 3D Brillouin zone and its projection onto the (001) surface. Blue and red dots mark the approximate positions of the Weyl points as predicted in Ref.~\cite{Soluyanov2015}. The gray line crossing $\overbar{\Gamma}$ and $\overbar{\text{W}}$ indicates the momentum cut investigated in the present TRARPES study. (c) ARPES intensity maps in $\overbar{\Gamma}$-$\overbar{\text{W}}$ direction before ($\Delta t = \SI{-1}{\pico\second}$) and after ($\Delta t = \SI{120}{\femto\second}$) excitation with \SI{827}{\nano\meter} pump pulses.}
		\label{fig:figure_01}
	\end{figure}

	With their first experimental observation in 2015, topological Weyl semimetals (WSMs) have attracted enormous attention~\cite{Lv2015,Yang2015,Xu2015,Xu2015b}. In these materials, long sought-after Weyl fermions are realized as quasi-particle excitations in condensed matter~\cite{Wan2011,Weng2015,Huang2015}. WSMs feature an unusual electronic structure with topologically protected crossing points in the bulk band structure, the so-called Weyl points. In the particular case of type-II WSMs, the Weyl points are located at touching points of electron and hole pockets close to the Fermi level $E_\text{F}$~\cite{Soluyanov2015,Sun2015,Chang2016}. The transition metal dichalcogenide \textit{Td}-\ce{WTe2} was the first material proposed to support such a scenario and band structure calculations using low-temperature lattice parameters predicted the existence of four pairs of Weyl points in the $\Gamma\text{XY}$ plane of the Brillouin zone [see Fig.~\ref{fig:figure_01}(b)]~\cite{Soluyanov2015}. Even though angle-resolved photomission spectroscopy (ARPES) studies did not succeed in observing Weyl points in this material so far~\cite{Bruno2016,Wang2016,Sanchez2016,Feng2016,Wu2016,DiSante2017,Zhang2017}, the observation of Weyl orbit related quantum oscillations and an anisotropic magnetoresistance~\cite{Li2017}, the detection of Weyl points via scanning tunneling spectroscopy~\cite{Lin2017}, and the observation of an anisotropic Adler-Bell-Jackiw anomaly~\cite{Lv2017} give strong experimental evidence that \textit{Td}-\ce{WTe2} is indeed a type-II WSM.
	
	The lack of an inversion center of the crystal structure is the prerequisite for a WSM phase in non-magnetic materials~\cite{Soluyanov2015}. Recently, an ultrafast and reversible way of manipulating the structural symmetry of \textit{Td}-\ce{WTe2} was demonstrated in a combined ultrafast electron diffraction (UED) and time-resolved second-harmonic generation (TRSHG) study~\cite{Sie2019}. Upon excitation with terahertz pump pulses, a coherent \SI{0.24}{\tera\hertz} interlayer shear excitation was observed that, at sufficiently high excitation densities, drives a structural phase transition from the non-centrosymmetric \textit{Td} ground state of the material [see Fig.~\ref{fig:figure_01}(a)] into a meta-stable centrosymmetric $1\textit{T}'(^*)$ phase~\cite{Sie2019}. Band structure calculations imply the periodic modulation of the Weyl point intra-pair separation upon excitation of the shear mode. Even more, the complete annihilation of the Weyl points is expected as the material undergoes the transition into the $1\textit{T}'(^*)$ phase. However, experimental data on how the shear mode affects the band structure of \textit{Td}-\ce{WTe2} is still lacking.
	
	Here, we present a time-resolved ARPES (TRARPES) study on the electronic structure response of \textit{Td}-\ce{WTe2} to the excitation of coherent phonons. Upon absorption of \SI{827}{\nano\meter} femtosecond laser pulses, we observe in the TRARPES data clear oscillations in photoemission (PE) intensity, band positions, and band widths. The comparison with Raman spectroscopy~\cite{Jiang2016,Kong2015,Song2016} and optical pump-probe spectroscopy results~\cite{Dai2015,He2016} allows assigning these oscillations to the excitation of five different A$_1$ optical phonon modes, with one of them being the interlayer shear mode mentioned above. An energy- and momentum-resolved Fourier transformation of the TRARPES data enables us to perform a phonon mode-selective analysis of the electronic structure response. We observe that the excitation of the interlayer shear mode periodically modulates occupied bands that are spin split due to the broken inversion symmetry of the crystal including a hole pocket that is directly involved in the formation of the Weyl points at low temperatures~\cite{Kim2017}. Even more, we observe that also the PE signal from the energy-momentum region of the Weyl points shows clear oscillations at the shear mode frequency. Although the presented experiments were performed at room temperature, for which theory excludes the presence of Weyl points~\cite{Bruno2016}, our results give direct experimental support for a coherent phonon mediated control of the electronic structure relevant for the Weyl physics in \textit{Td}-\ce{WTe2}.            
	
	\section*{Results}
	
	\begin{figure}
		\centering
		\includegraphics[width=1\linewidth]{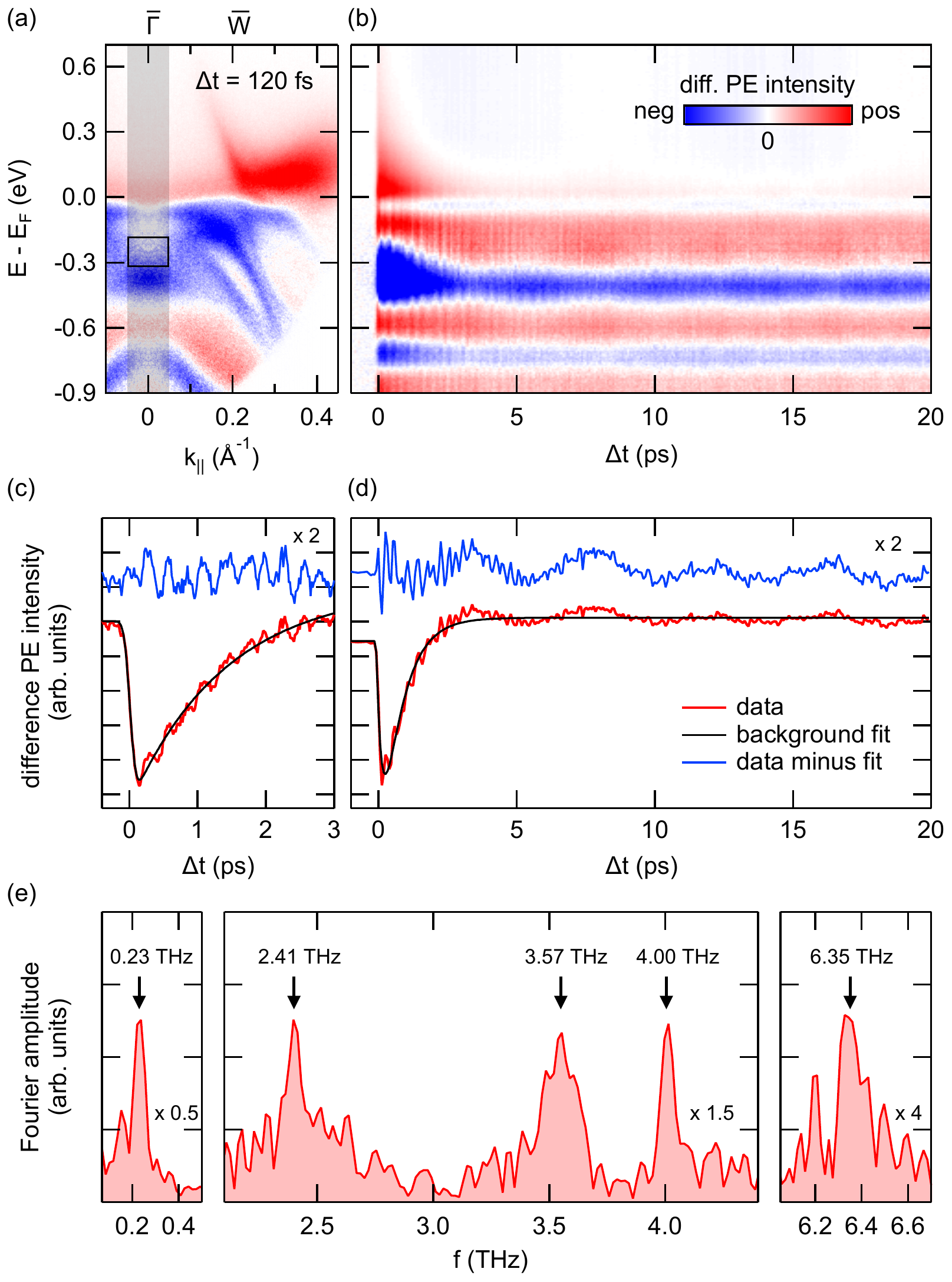}
		\caption{TRARPES results of \textit{Td}-\ce{WTe2} along the $\overbar{\Gamma}$-$\overbar{\text{W}}$ direction. (a) ARPES difference intensity map derived from the TRARPES data at $\Delta t = \SI{120}{\femto\second}$ and $\Delta t = \SI{-1}{\pico\second}$ shown in Fig.~\ref{fig:figure_01}(c). (b) Transient difference EDCs derived from the ARPES difference intensity map by momentum integration of the gray-shaded area in (a). (c)-(d) PE intensity transients of the integration region marked by the black box in (a) before (red) and after (blue) background subtraction (Supplementary Note 3). Blue curves are offset and scaled by a factor of $2$ for clarity. (e) Fourier amplitude spectrum of the intensity transient shown in (d). Note the different scaling of the data.}
		\label{fig:figure_02}
	\end{figure}
	
	Figure~\ref{fig:figure_01}(c) shows TRARPES data of  \textit{Td}-\ce{WTe2} recorded before the optical excitation ($\Delta t = \SI{-1}{\pico\second}$) in comparison with data recorded at a pump-probe delay of $\Delta t = \SI{120}{\femto\second}$. Overall, the spectra are consistent with ARPES and TRARPES data recorded at a similar probe photon energy of $h\nu \approx \SI{6}{\electronvolt}$~\cite{Bruno2016, Caputo2018, Wu2016}. In agreement with previous TRARPES studies, we observe the transient population of an electron pocket above $E_\text{F}$ in response to the near-infrared (NIR) excitation~\cite{Crepaldi2017,Caputo2018,Das2019}. A difference intensity map of the two ARPES spectra [see Fig.~\ref{fig:figure_02}(a)] emphasizes the presence of spectral changes also below $E_\text{F}$. Part of the observed transient reduction of spectral weight (blue areas) results from the depopulation of the occupied bands due to the absorption process. However, the observation of an increase in spectral weight in some regions below $E_\text{F}$ (red areas) hints also to the presence of transient band renormalization processes. 
	
		\begin{figure}
		\centering
		\includegraphics[width=1\linewidth]{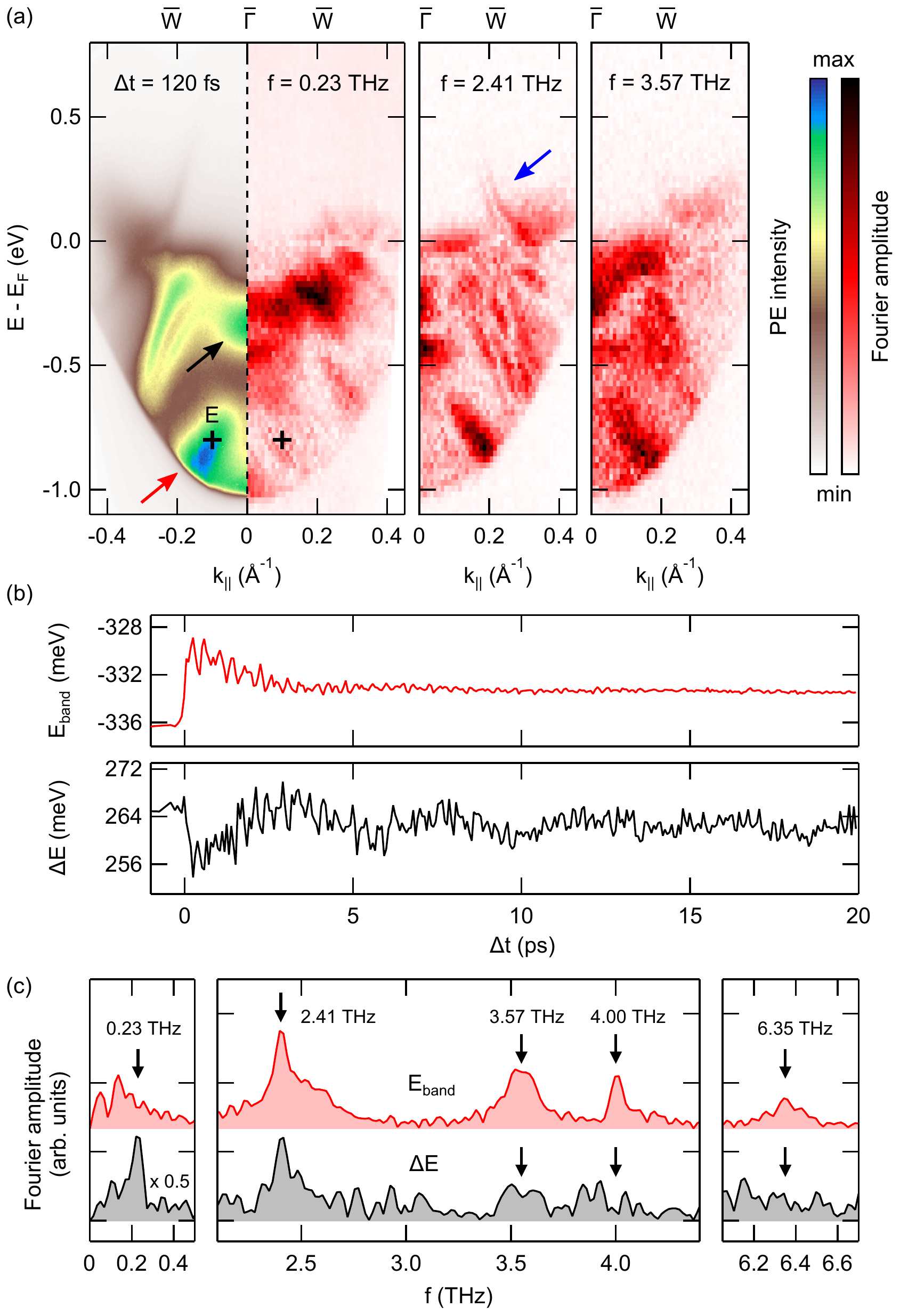}
		\caption{Phonon mode-selective Fourier amplitude analysis of the TRARPES data. (a) ARPES intensity map at $\Delta t = \SI{120}{\femto\second}$ in comparison to Fourier maps of different phonon frequencies. Fourier maps were generated from ARPES intensity maps that were binned over regions of $\SI{23}{\milli\electronvolt} \times \SI{0.01}{\per\angstrom}$ in order to improve the signal statistics. Arrows indicate energy-momentum regions that are discussed in more detail in the main text. The cross marks the position of ROI E. Fourier maps of the \SI{4.00}{\tera\hertz} mode and the \SI{6.35}{\tera\hertz} mode are shown in Supplementary Fig. S.5. (b) Transient peak energy $E_{\text{band}}$ and peak width $\Delta E$ (FWHM) of the ARPES signal marked by the black arrow in (a). The data result from Gaussian fits of the transient EDCs at $\overbar{\Gamma}$. PE intensity was integrated over a momentum window of \SI{0.05}{\per\angstrom}. (c) Fourier amplitude spectra of the transients shown in (b). The peaks below \SI{0.23}{\tera\hertz} can not be assigned to any phonon modes in \textit{Td}-\ce{WTe2}. Their origin is discussed in Supplementary Note 2.}
		\label{fig:figure_03}
	\end{figure}
	
	Figure~\ref{fig:figure_02}(b) shows transient difference energy distribution curves (EDCs) around the $\overbar{\Gamma}$ point as a function of $\Delta t$. Above $E_\text{F}$, the data clearly reveals the presence of an excited carrier population that decays on a timescale of several picoseconds. Furthermore, part of the transient spectral changes are periodically modulated indicative for the excitation of at least two coherent phonon modes exhibiting oscillation periods in the few \SI{100}{\femto\second} and few \SI{}{\pico\second} range, respectively. Significant long-lived spectral changes survive the damping of the coherent phonon modes and show barely any changes even at delays of \SI{400}{\pico\second} (Supplementary Note 1). Our observations qualitatively resemble the findings of previous time-resolved studies of \textit{Td}-\ce{WTe2}: Electronic excitation and relaxation processes were studied in detail using TRARPES~\cite{Crepaldi2017, Caputo2018, Das2019} and time-resolved reflectivity (TRR) measurements~\cite{Dai2015}. The excitation of coherent phonons was observed in TRR~\cite{Dai2015,He2016} and TRSHG experiments~\cite{Sie2019} as well as via UED~\cite{Sie2019}, but, notably, in none of the past TRARPES studies~\cite{Crepaldi2017,Caputo2018,Das2019}. Also the observed long-lived spectral changes are compatible with the findings of different other studies~\cite{Sie2019,Dai2015}. In the following, we will exclusively focus on the analysis of the coherent phonon oscillations and their impact on the electronic structure of \textit{Td}-\ce{WTe2}.
	
\begin{figure}
		\centering
		\includegraphics[width=1\linewidth]{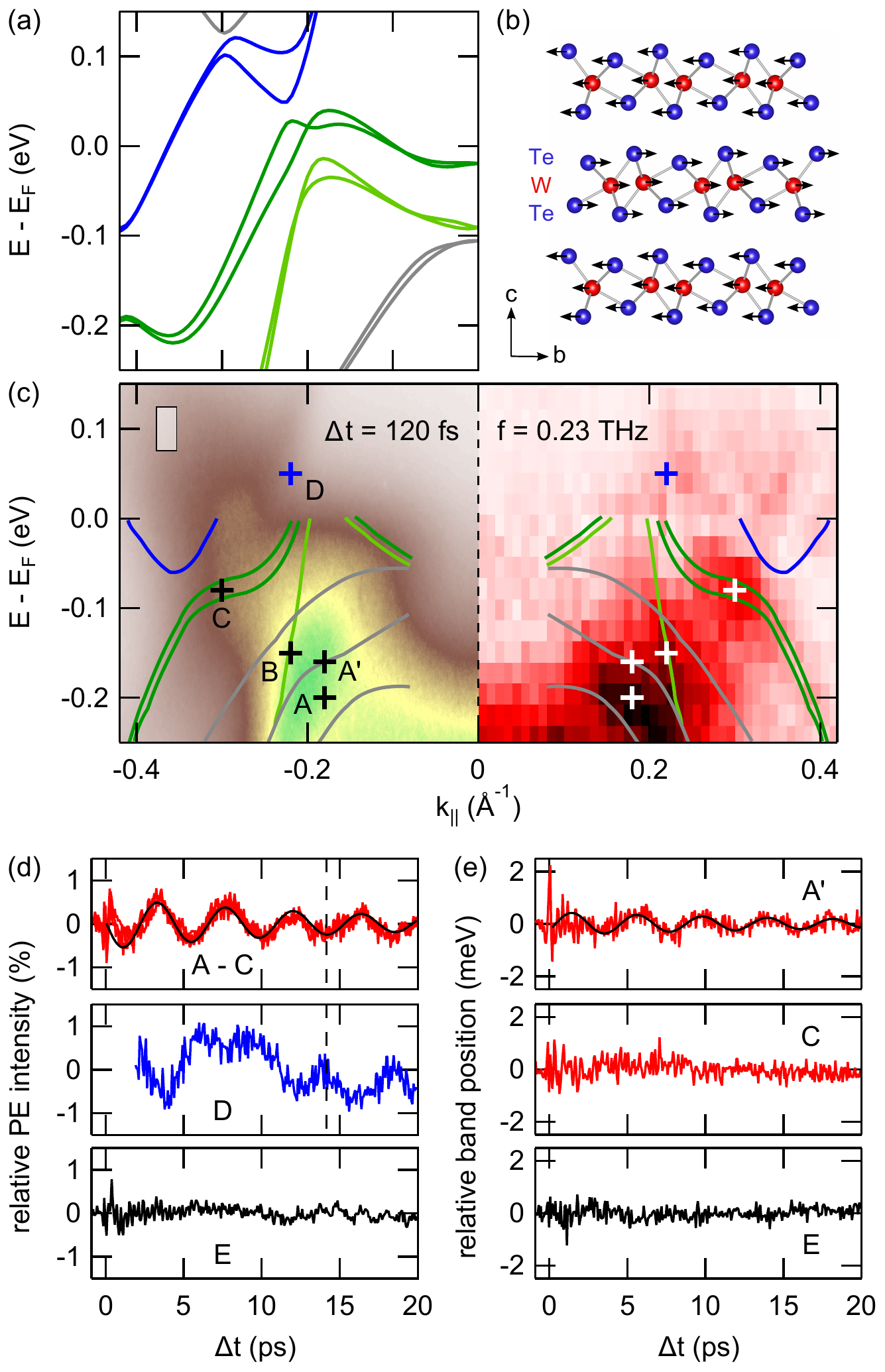}
		\caption{Response of the electronic structure of \textit{Td}-\ce{WTe2} to the excitation of the \SI{0.23}{\tera\hertz} interlayer shear mode. (a) Calculated bulk band structure along the $\Gamma$-X direction. (b) Schematic illustration of the atomic motion upon excitation of the interlayer shear mode~\cite{Chen2016}. (c) Close-ups of the ARPES intensity map near $E_\text{F}$ at $\Delta t = \SI{120}{\femto\second}$ (left) and the corresponding \SI{0.23}{\tera\hertz} Fourier map (right). The data is overlaid with experimental band dispersions from Ref.~\cite{Bruno2016}. Surface bands were omitted for the sake of clarity. Crosses mark positions of the ROI A-D and A$'$. The Fourier map was generated from ARPES intensity maps that were binned over regions of $\SI{23}{\milli\electronvolt} \times \SI{0.01}{\per\angstrom}$. (d) PE intensity transients of ROI A-E. The transients of ROI A-C (top graph) can barely be discerned and are all displayed in red. The size of the integration area used to generate the transients is indicated by the gray shaded box in the ARPES data shown in (c). (e) Peak shift analysis of ROI A$'$, C, and reference region E marked in Fig.~\ref{fig:figure_03}(a). All transients in (d) and (e) are background-corrected (Supplementary Note 3). The black lines in the top graphs of (d) and (e) are fits of a damped sinusoidal function.}
		\label{fig:figure_04}
	\end{figure}

	Figures~\ref{fig:figure_02}(c) and \ref{fig:figure_02}(d) show PE intensity transients around $\overbar{\Gamma}$ and $E - E_\text{F} = \SI{-0.25}{\electronvolt}$. The data were taken with different sampling rates (\SI{60}{\tera\hertz} and \SI{15}{\tera\hertz}) and cover different delay ranges (\SI{3}{\pico\second} and \SI{20}{\pico\second}) to separately illustrate both the high- and low-frequency contributions to the PE intensity modulations (Supplementary Note 2). 
	Raw data (red lines) were fitted with an exponential model function (Supplementary Note 3) to account for the carrier population dynamics (black lines) and to extract the pure oscillatory part of the signals (blue lines). The beating of the \SI{3}{\pico\second} range data clearly reveals the presence of more than one high-frequency mode. Additionally, a well-separated low-frequency modulation can be identified in the \SI{20}{\pico\second} range data. Figure~\ref{fig:figure_02}(e) shows a Fourier amplitude spectrum of the \SI{20}{\pico\second} range intensity transient. Overall, we find main peaks at \SI{0.23}{\tera\hertz}, \SI{2.41}{\tera\hertz}, \SI{3.57}{\tera\hertz}, \SI{4.00}{\tera\hertz}, and \SI{6.35}{\tera\hertz} indicative for the excitation of at least five different coherent phonon modes. A comparison with results of Raman studies of \ce{WTe2} and \ce{MoTe2} allows assigning all frequencies to A$_1$ optical phonon modes~\cite{Jiang2016,Kong2015,Song2016,Chen2016} that belong to the group of \textit{m}-modes with the atoms vibrating in the $bc$-mirror plane of the unit cell~\cite{Chen2016}. Of particular relevance is the assignment of the \SI{0.23}{\tera\hertz} mode, which is responsible for the distinct long-periodic spectral modulations visible in Fig.~\ref{fig:figure_02}(b) and \ref{fig:figure_02}(d): In agreement with UED, TRSHG and TRR results~\cite{Sie2019,Dai2015,He2016}, and predictions from Raman studies~\cite{Jiang2016,Kong2015,Song2016,Chen2016}, we assign this frequency to the low-energy optical phonon interlayer shear mode along the $b$ axis [see Fig.~\ref{fig:figure_04}(b)]. The excitation of this mode periodically drives \ce{WTe2} from its non-centrosymmetric \textit{Td} structure towards a centrosymmetric $1\textit{T}'(^*)$ structure. It is therefore expected to periodically modulate the spin splitting of bands~\cite{Kim2017} as well as the Weyl point intra-pair separation in this material~\cite{Sie2019}.

	Figure~\ref{fig:figure_03}(a) shows the results of an energy- and momentum-resolved Fourier analysis of the TRARPES data (Supplementary Note 4) illustrating the phonon mode-selectivity of the electronic structure response. The figure separately displays energy-momentum maps of the Fourier amplitudes of the \SI{0.23}{\tera\hertz}, \SI{2.41}{\tera\hertz}, and \SI{3.57}{\tera\hertz} coherent phonon excitations in comparison to the ARPES intensity map at $\Delta t = \SI{120}{\femto\second}$. The data clearly illustrates the band-selectivity of the phonon excitations, as seen for instance for the band marked by the red arrow. A more detailed analysis shows, furthermore, that even in the case that the same band is affected by several modes, its response can be quite different. This is illustrated by band energy ($E_{\text{band}}$) and band width ($\Delta E$) transients [Fig. \ref{fig:figure_03}(b)] and their Fourier amplitude spectra [Fig. \ref{fig:figure_03}(c)] for the feature in the TRARPES map centered at $\overbar{\Gamma}$ at an energy $E-E_\text{F} \approx \SI{-0.3}{\electronvolt}$ marked by the black arrow. The excitation of the four high-frequency modes results in small but clearly detectable shifts in $E_{\text{band}}$, which becomes particularly evident in the Fourier amplitude spectrum of the band energy transient. For the excitation of the \SI{0.23}{\tera\hertz} interlayer shear mode, such a band shift, if present at all, stays below the detection limit. In contrast, a finite and - also in comparison to the other modes – now significant contribution of the shear mode to the broadening of the band is evident from the $\Delta E$ transient as well as from the \SI{0.23}{\tera\hertz} peak in the corresponding Fourier amplitude spectrum. It is finally noteworthy that the Fourier analysis of the TRARPES data not only shows a finite signal amplitude below $E_\text{F}$ but also at energies above $E_\text{F}$ up to $E-E_\text{F} \approx \SI{0.3}{\electronvolt}$ (see blue arrow). This sensitivity partly results from the laser excitation, which transiently heats up the electron gas so that states above $E_\text{F}$ are substantially populated even on timescales of several \SI{10}{\pico\second}.
	
	In comparison to the ARPES intensity maps, the Fourier maps exhibit additional fine structures showing spectral details even below the \SI{40}{\milli\electronvolt} energy resolution of the experiment. The fine structure is particularly striking in the \SI{2.41}{\tera\hertz} Fourier map. A comparison with experimental data of a low temperature, high-resolution ARPES study \cite{Bruno2016} shows that the fine structure resembles in large part the band structure of \textit{Td}-\ce{WTe2}. We account non-linearities in the dynamical signal response to the coherent phonon excitation being responsible for this resolution enhancement. More details are given in Supplementary Note 5.
	
	Apart from the mode-selective amplitude analysis, we also performed a mode-selective phase analysis of the transient peak energies (Supplementary Note 6). The four high-frequency oscillations all show a cosinusoidal behavior with respect to time zero of the experiment implying a displacive excitation of the coherent phonons~\cite{Zeiger1992}. For the \SI{0.23}{\tera\hertz} mode, we observe, in contrast, a sinusoidal modulation of the peak energies, in agreement with UED and TRSHG results using THz and NIR excitation pulses, respectively~\cite{Sie2019}. Typically, such a response is associated with an impulsive Raman stimulated excitation mechanism~\cite{Garrett1996}. It should be added that for the excitation of the interlayer shear mode in \textit{Td}-\ce{WTe2} using \SI{}{\tera\hertz} pulses a field driven photo-doping process was alternatively suggested to explain the observed sinusoidal response~\cite{Sie2019}.  
	
	In the further discussion, we will focus on	the excitation of the \SI{0.23}{\tera\hertz} interlayer shear mode periodically modulating the \ce{WTe2} crystalline \textit{Td} structure towards a centrosymmetric $1\textit{T}'(^*)$ structure~\cite{Sie2019}. Based on values given in Ref.~\cite{Sie2019} and under consideration of a TRR study at \SI{800}{\nano\meter} excitation~\cite{He2016}, we estimate the shear displacement amplitude at the excitation fluence applied in our experiment to be in the order of \SI{1}{\pico\meter} (Supplementary Note 7).   
	
	Figure~\ref{fig:figure_04}(c) shows close-ups of the ARPES intensity map at $\Delta t = \SI{120}{\femto\second}$ and the \SI{0.23}{\tera\hertz} Fourier map of Fig.~\ref{fig:figure_03}(a). The data is overlaid with the bulk band structure of \textit{Td}-\ce{WTe2} along the $\overbar{\Gamma}$-$\overbar{\text{X}}$ direction experimentally determined in a high-resolution ARPES study at $h\nu = \SI{6.01}{\electronvolt}$~\cite{Bruno2016}. The color coding was chosen to discriminate between an electron pocket (blue), two hole pockets (green), and further bulk bands (gray). Calculated band structure data along $\Gamma$-$\text{X}$ is shown for comparison in Fig.~\ref{fig:figure_04}(a). Due to thermal broadening and the limited energy resolution of our experiment	in combination with strong variations in the spectral weight among the different bands~\cite{Bruno2016}, it is difficult to discern the individual bands in the ARPES intensity map. However, even though broadened, the dispersing amplitude maxima in the Fourier map match part of the experimental band structure data strikingly well. The main amplitude maximum at $k_\parallel \approx\SI{0.2}{\per\angstrom}$ shows a branching [see also Fig.~\ref{fig:figure_03}(a)] that follows the dispersion of the lower hole pocket and the two gray colored lower bulk bands, respectively. The data also reproduces the dispersion of the upper hole pocket up to energies close to $E_\text{F}$ with a distinct amplitude maximum in the plateau region at $k_\parallel \approx\SI{0.3}{\per\angstrom}$. Above $E_\text{F}$, we observe two weak but distinct local amplitude maxima at $k_\parallel \approx\SI{0.22}{\per\angstrom}$ and $k_\parallel \approx\SI{0.34}{\per\angstrom}$ with the former one being approximately located at the surface projection of the Weyl points marked by the blue cross in Fig.~\ref{fig:figure_04}(c)~\cite{Soluyanov2015, Lin2017}.
	
	Within the energy momentum region probed in our experiments, different ARPES studies reported also on the observation of surface states \cite{Bruno2016, Wang2016, Sanchez2016, Wu2016}. Owing to the limited energy resolution and the fact that our experiments were performed at room temperature, no distinct signatures of these surface states could be observed in our ARPES intensity maps. A comparison with the experimental surface state data of Ref. \cite{Bruno2016} showed, furthermore, that also the \SI{0.23}{\tera\hertz} Fourier map yields no clear indication for a response of the surface states to the excitation of the shear mode. In Fig. \ref{fig:figure_04}(c) we therefore omitted the inclusion of the surface bands for the sake of clarity. A comparison including surface bands is, however, provided in Supplementary Note 8.
	
	For the further analysis, we selected four regions of interest (ROI) in the Fourier maps at positions indicated by the crosses in Fig.~\ref{fig:figure_04}(c). ROI A, B, and C are located at Fourier amplitude maxima. ROI B and C are furthermore intersected by the lower and upper hole pocket, respectively, whereas ROI A is positioned in between the two lowest bulk bands. ROI D is centered on the surface projection of the Weyl points above $E_\text{F}$. For reference, we selected an additional ROI E [see Fig.~\ref{fig:figure_03}(a)] at a local intensity maximum in the ARPES intensity maps that, however, shows a negligible amplitude in the \SI{0.23}{\tera\hertz} Fourier map. Where possible, we evaluated for the different ROI the transient evolution of the PE intensity, peak energy, and peak width in response to the excitation of the interlayer shear mode. For all ROI, we chose a signal integration area of $\SI{50}{\milli\electronvolt} \times \SI{0.025}{\per\angstrom}$.
	
	Results of the PE intensity analysis are summarized in Fig.~\ref{fig:figure_04}(d). For ROI A, B, and C, we observe clear periodic modulations at the frequency of the shear mode. Relative amplitude and phase of the PE intensity transients of the ROI match each other extremely well [see top graph of Fig.~\ref{fig:figure_04}(d)] and can be described by a damped $\pi$-shifted sinusoidal function with a relative amplitude in the order of \SI{1}{\percent} (see full black line). Also the PE intensity of ROI D, which probes the Weyl point region, shows clear oscillations at \SI{0.23}{\tera\hertz} in spite of some signal distortions at $\Delta t \approx \SI{7.5}{\pico\second}$ resulting from limitations of the signal background subtraction (Supplementary Note 3). Notably, the oscillatory response of ROI D shows a $\pi$-phase shift with respect to ROI A, B, and C as indicated by the black dashed line and also with respect to the second amplitude maximum above $E_\text{F}$ at $k_\parallel \approx\SI{0.34}{\per\angstrom}$ (Supplementary Note 9). Contrary to ROI A-D, the transient PE intensity signal of ROI E exhibits no evidence for a response to the shear mode as expected from the vanishing signal in the Fourier map.
	
	A peak shift analysis is only possible for ROI that are located at signal maxima in the ARPES intensity maps. This only applies to ROI C and E. To account for potential spectral shifts affecting ROI A, we selected a nearby PE intensity maximum (ROI A$'$) separated by $\approx \SI{40}{\milli\electronvolt}$ and intersected by the center bulk band. Distinct PE intensity maxima that could be associated with ROI B and D could not be identified. Results of the peak shift analysis are shown in Fig.~\ref{fig:figure_04}(e). A distinct shift in the peak energy at the shear mode frequency is only observed for ROI A$'$. The oscillation follows a damped sinusoidal function with an amplitude of approximately \SI{1}{\milli\electronvolt}. The data reveals no detectable peak shift at the shear mode frequency for ROI C and E. The analysis of the peak widths for ROI A$'$, C, and E did not yield any resolvable changes at the shear mode frequency. However, a detailed Fourier analysis of ROI C shows indirect evidence for subtle changes in the peak width as will be discussed below. 
	
	\section*{Discussion}
	
	Ref. \cite{Kim2017} reports on the evolution of the band structure of \ce{WTe2} as the crystalline structure undergoes a complete transition from the \textit{Td} phase to the centrosymmetric $1\textit{T}'$ phase upon external charge doping. The calculations show that the changes in the band structure predominantly result from changes in the interlayer interaction that is in an analog manner periodically modulated upon excitation of the \SI{0.23}{\tera\hertz} interlayer shear mode following the photoexcitation with ultrashort NIR pulses. More specifically, such a transition is expected to reduce and finally annihilate the spin splitting of the different bands near $E_\text{F}$ as the inversion symmetry of the crystalline structure is recovered. These findings imply that the oscillatory part of the PE intensity transients from the two hole pockets (ROI B and C) results from a periodic modulation of the spin splitting due to the shear mode excitation. Notably, for the upper hole pocket the Fourier amplitude is indeed maximum in the region at which the experimental band structure data implies the largest spin splitting and also for the lower hole pocket, band structure calculations predict a considerable spin splitting [see Fig.~\ref{fig:figure_04}(a)].  
	
	\begin{figure}
		\centering
		\includegraphics[width=1\linewidth]{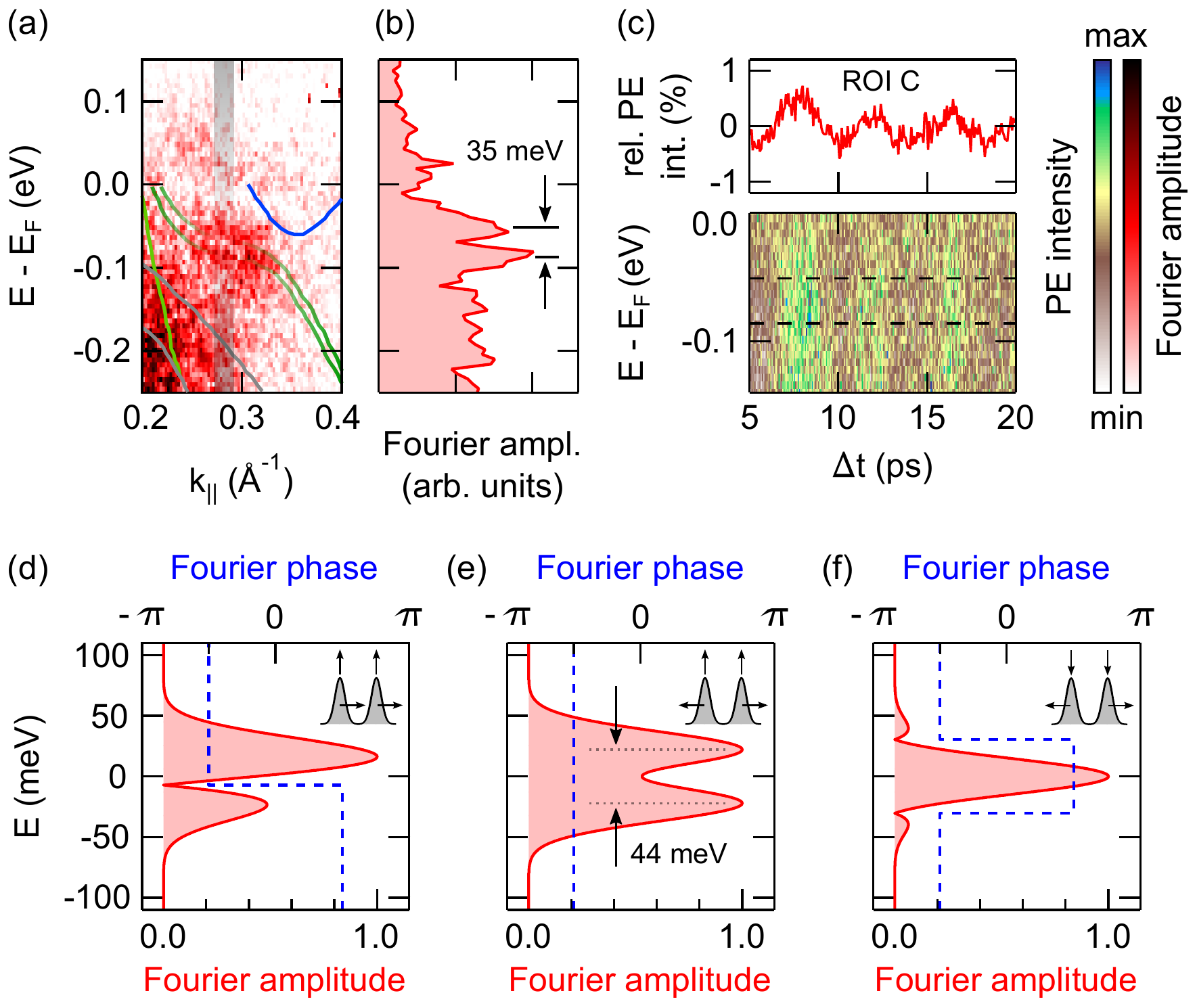}
		\caption{Fourier amplitude modulation near ROI C. (a) Close-up of the \SI{0.23}{\tera\hertz} Fourier map near ROI C. The data is overlaid with experimental band dispersions from Ref.~\cite{Bruno2016}. For better visibility of the Fourier amplitude modulation near ROI C, the spin split upper hole bands are kept semi-transparent in this region. The energy-momentum cut displayed in (b) is indicated. The Fourier map was generated from ARPES intensity maps that were binned over regions of $\SI{6}{\milli\electronvolt} \times \SI{0.0025}{\per\angstrom}$. (b) Fourier amplitude as a function of binding energy extracted from the Fourier map along the energy-momentum cut indicated in (a). The separation of the amplitude maxima near ROI C is indicated. (c) Color-coded plot of experimental EDCs along the evaluated energy-momentum cut near ROI C as a function of $\Delta t$. The dashed lines mark the binding energies of the two Fourier amplitude maxima indicated in (b). The top graph displays for comparison the PE intensity transient of ROI C. (d)-(f) Fourier amplitude and Fourier phase as a function of binding energy resulting from the simulations of (d) scenario (i), (e) scenario (ii), and (f) scenario (iii). The pictograms in the graphs schematically illustrate the considered scenarios. Details are given in the text and in Supplementary Note 10. The separation of the Fourier amplitude maxima for scenario (ii) is indicated.}
		\label{fig:figure_05}
	\end{figure}

	A more detailed inspection of the Fourier amplitude signal near ROI C further confirms that the shear mode excitation modulates the spin splitting. Figure \ref{fig:figure_05}(a) shows the \SI{0.23}{\tera\hertz} Fourier map near ROI C. In comparison to Figs.~\ref{fig:figure_03}(a) and \ref{fig:figure_04}(c), the original ARPES data were in this case binned over smaller energy-momentum regions for further data processing so that some more details become visible at the cost of signal statistics. The pronounced maximum in the Fourier map at ROI C appears now split by a weak but distinct amplitude minimum that follows the experimental band structure data from Ref.~\cite{Bruno2016} strikingly well. A constant momentum cut across the Fourier amplitude maximum [Fig. \ref{fig:figure_05}(b)] confirms this splitting and yields a separation of the resulting two  Fourier amplitude maxima by $\approx\SI{35}{\milli\electronvolt}$. To gain insight into the origin of this Fourier amplitude modulation, we performed simulations mimicking the following three potential scenarios: (i) a rigid shift of the spin-split bands accompanying an overall PE intensity oscillation, (ii) a modulation of the spin splitting that is in phase with a PE intensity oscillation, i.e., a decrease (increase) of the spin splitting is accompanied by a decrease (increase) of the PE intensity, and (iii) a modulation of the spin splitting that is out of phase with a PE intensity oscillation. Details on the simulations, including the choice of parameters, are described in Supplementary Note 10.
	
	Figures \ref{fig:figure_05}(d)-(f) show the resulting Fourier amplitudes and phases as a function of energy for the three scenarios. Whereas the amplitude modulation calculated for scenario (iii) clearly deviates from the experimental results, a reasonable match is observed for scenario (i) and scenario (ii). However, inspection of the phase behavior shows that only scenario (ii) can consistently reproduce the experimental data:  Figure \ref{fig:figure_05}(c) depicts a color-coded plot of experimental EDCs along the evaluated energy-momentum cut near ROI C as a function of $\Delta t$. The data reveal an in-phase response of the PE signal independent of binding energy across the two Fourier amplitude maxima (see dashed lines), in agreement with the constant phase expected for scenario (ii) and contrary to the $\pi$-phase shift right at the Fourier amplitude minimum expected for scenario (i). Notably, the simulation reproduces the experimental amplitude modulation as well as the separation of the two amplitude maxima also quantitatively strikingly well and implies that in our experiment the spin splitting is periodically modulated by the excitation of the shear mode at an amplitude in the order of \SI{1}{\milli\electronvolt} (see Supplemental Note 10). It is finally interesting to also consider the initial phase of the oscillation in the spin splitting. The excitation of the shear mode results in an initial reduction of the PE intensity in ROI C [see Fig. \ref{fig:figure_04}(d)]. For the case of scenario (ii), this implies an initial reduction of the spin splitting, i.e., an initial shear motion towards the centrosymmetric $1\textit{T}'(^*)$ structure, in perfect agreement with the observations reported in Ref. \cite{Sie2019}.
	
	Changes in the spin splitting should also affect the response in the Weyl point area (ROI D). However, in this case the situation is more complex as the area covers at the same time the signal from the close-lying and spin-split electron pocket and upper hole pocket [see Fig.~\ref{fig:figure_04}(a)], which at low temperatures give rise to the formation of the Weyl points. A reduction (increase) in the spin splitting upon excitation of the shear mode will at the same time increase (reduce) the separation between electron pocket and hole pocket \cite{Kim2017}. The anti-phase behavior of the latter process with respect to the spin splitting may explain why the PE signal from the Weyl point area is $\pi$-shifted in comparison to all other ROI.

	Ref.~\cite{Kim2017} finally provides also information on how the highest of the three bulk bands will be affected by a shear motion along the $b$ axis. Due to the much smaller spin splitting of this band in comparison with the electron and hole pockets, the calculation predicts only very subtle changes even for a full phase transition. In the Fourier map, this band is indeed the only band which lacks of a clear signal along its band dispersion. The two other bulk bands are in contrast not considered in the calculations in Ref.~\cite{Kim2017}. However, band structure calculations imply that these two bands show a distinct dispersion along the $\Gamma$-Z direction (Supplementary Note 11), i.e., in $c$-direction [see Fig.~\ref{fig:figure_01}]. It is therefore not surprising that particularly these bands become affected by a shear displacement among neighboring \ce{WTe2} layers giving rise to rather large amplitudes in the Fourier map. \\
	
    The excitation of coherent phonons provides unique opportunities for the study and coherent control of structural, electronic, and magnetic properties of solids~\cite{Kim2012,Yusupov2010,Gerber2017}. Time- and angle-resolved photoemission spectroscopy is in this context the most direct instrument to map the electronic structure response at the required energy- and momentum sensitivity ~\cite{Schmitt2008,Papalazarou2012,Hellmann2012}. The phonon mode-resolved Fourier maps introduced in the present work allow in a very direct and intuitive manner for an electron-band selective view onto electron-phonon interaction processes including even electronic states above $E_\text{F}$. The striking differences observed among the Fourier maps of \textit{Td}-\ce{WTe2} emphasize the band-selectivity of coherent phonon excitation processes. Remarkably, in the Fourier map representation, the nonlinear signal response to the excitation of coherent phonons can substantially enhance the spectral resolution uncovering spectral details not visible in the ARPES spectra.
    
    Our results reveal, furthermore, that the excitation of a low-frequency interlayer shear mode periodically modulates the spin splitting of bands, a spectral signature that is closely linked to the broken inversion symmetry of the crystalline lattice. In addition, the data prove that the excitation of the shear mode affects the electronic structure in the energy-momentum area comprising the Weyl points in \textit{Td}-\ce{WTe2} at low temperatures. Overall, the presented experimental results strongly support the relevance of the shear mode excitation for the control of the specific Weyl physics in this material as recently predicted in Ref. \cite{Sie2019}. 
    
    \section*{Methods}
    
    \textbf{Sample synthesis.} High-quality \textit{Td}-\ce{WTe2} crystals were grown using a chemical vapour transport technique. Stoichiometric tungsten powder (\SI{99.9}{\percent}) and tellurium powder (\SI{99.99}{\percent}) were ground together and loaded into a quartz tube with a small amount of the transport agent \ce{TeBr4}. All weighing and mixing was carried out in a glove box. The tube was sealed under vacuum and placed in a two-zone furnace. The hot zone and the cold zone were maintained for one week at a constant temperature of \SI{800}{\celsius} and \SI{700}{\celsius}, respectively. \\
    
    \textbf{TRARPES experiments.} For the TRARPES experiments, we used two non-collinear optical parametric amplifiers (NOPAs) which are pumped by the second harmonic of a chirped pulse amplifier. One of the NOPAs delivers \SI{1.5}{\electronvolt} (\SI{827}{\nano\meter}), \SI{30}{\femto\second}, p-polarized pump pulses with an incident fluence of \SI{110}{\micro\joule\per\square\centi\meter} on the sample. The \SI{840}{\nano\meter} output of the second NOPA system is used to generate \SI{5.9}{\electronvolt} (\SI{210}{\nano\meter}), \SI{95}{\femto\second}, s-polarized probe pulses by sequential frequency doubling. Cross-correlation measurements at the sample position yielded $\approx \SI{100}{\femto\second}$ FWHM (Supplementary Note 2).
    Right before the pump-probe experiments, the \textit{Td}-\ce{WTe2} crystals were cleaved under ultra-high vacuum conditions. The samples were aligned by low-energy electron diffraction in a direction crossing $\overbar{\Gamma}$ and the projection $\overbar{\text{W}}$ of the predicted position of two neighboring Weyl points onto the (001) surface [see Fig.~\ref{fig:figure_01}(b)]. ARPES spectra were recorded using a hemispherical analyzer at a total energy resolution of \SI{40}{\milli\electronvolt}. All experiments were performed at room temperature at a pressure of \SI{2e-10}{\milli\bar}. \\

    \textbf{Band structure calculations.} The electronic bulk band structure was calculated by \textit{ab-initio} calculation based on density functional theory (DFT) with Projector augmented-wave (PAW) method \cite{Bloechl1994} as implemented in the Vienna \textit{Ab-initio} Simulation Package (VASP) \cite{Kresse1996}. The exchange and correlation energies were considered on the level of the generalized gradient approximation (GGA) with a Perdew-Burke-Ernzerhof (PBE) functional \cite{Perdew1996}. The energy cutoff was set to be \SI{350}{\electronvolt} for the plane wave basis. We included the van der Waals corrections via a pair-wise force field of the Grimme method \cite{Grimme2006}. The experimental lattice constants from Ref. \cite{Mar1992} were used in all the calculations.
    
    \section*{Data availability}
    
    All data that support the findings of this study are available from the corresponding author upon reasonable request.
   
	\bibliography{MyCollection}
	
	\section*{Acknowledgments}
    This work was supported by the German Research Foundation (DFG) through projects INST 257/419-1 and INST 257/442-1 and by the National Natural Science Foundation of China (Grant No. 11774190).
  
    \section*{Author Contributions}
    P.H., S.J. and M.B. conceived the experiments. P.H., S.J., and H.E. carried out TRARPES measurements. P.H. performed the data analysis. Y.Q. synthesized and characterized the \ce{WTe2} single crystals. Y.S. performed \textit{ab-initio} calculations. P.H. and M.B. co-wrote the paper. All authors contributed to the scientific planning and discussions.
    
    \section*{Additional Information}
	\textbf{Competing interests:} The authors declare no competing financial interests.
	
\end{document}


\title{Supplemental Material for \\
        "Mode-resolved reciprocal space mapping of electron-phonon interaction in the Weyl semimetal candidate \textit{Td}-\ce{WTe2}"}

 \author{Petra Hein}

    \affiliation{Institute of Experimental and Applied Physics, University of Kiel, Leibnizstr. 19, D-24118 Kiel, Germany}
    \author{Stephan Jauernik}
    \affiliation{Institute of Experimental and Applied Physics, University of Kiel, Leibnizstr. 19, D-24118 Kiel, Germany}
    \author{Hermann Erk}
    \affiliation{Institute of Experimental and Applied Physics, University of Kiel, Leibnizstr. 19, D-24118 Kiel, Germany}

    \author{Lexian Yang}
    \affiliation{State Key Laboratory of Low Dimensional Quantum Physics, Collaborative Innovation Center of Quantum Matter and Department of Physics, Tsinghua University, Beijing 100084, P. R. China}

    \author{Yanpeng Qi}
    \affiliation{School of Physical Science and Technology, ShanghaiTech University, Shanghai 201210, China}
    \affiliation{Max Planck Institute for Chemical Physics of Solids, N\"othnitzer Str. 40, D-01187 Dresden, Germany}

    \author{Yan Sun}
    \affiliation{Max Planck Institute for Chemical Physics of Solids, N\"othnitzer Str. 40, D-01187 Dresden, Germany}

    \author{Claudia Felser}
    \affiliation{Max Planck Institute for Chemical Physics of Solids, N\"othnitzer Str. 40, D-01187 Dresden, Germany}

    \author{Michael Bauer}
    \affiliation{Institute of Experimental and Applied Physics, University of Kiel, Leibnizstr. 19, D-24118 Kiel, Germany}

    \date{\today}

    \maketitle

\section*{Supplementary Note 1: Long-lived spectral changes}

\begin{figure}
    \centering
\includegraphics[width=0.8\linewidth]{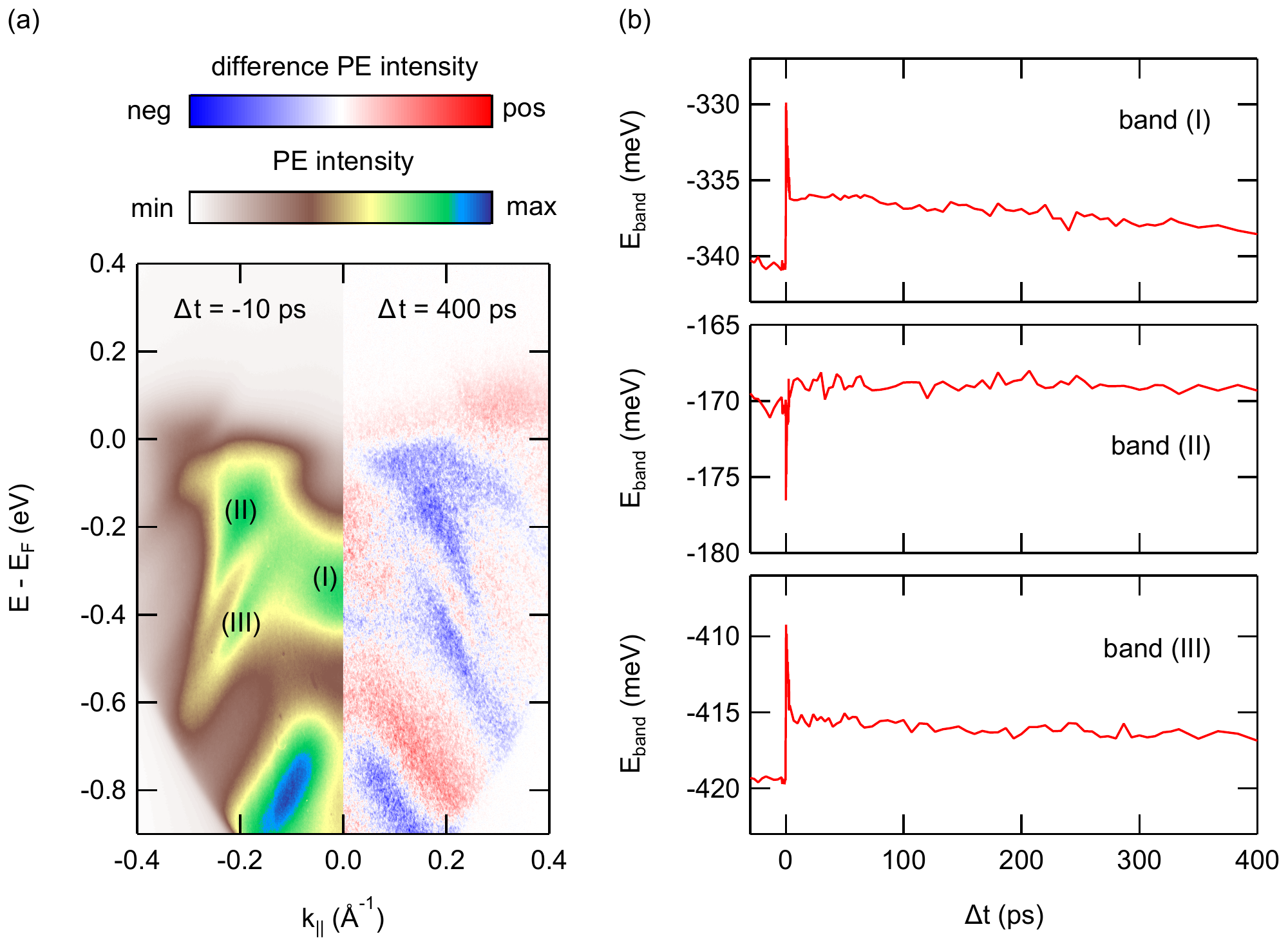}
    \caption{Long-lived spectral changes in the TRARPES data of \textit{Td}-\ce{WTe2}. (a) ARPES intensity map recorded at $\Delta t = -\SI{10}{\pico\second}$ in comparison to a difference intensity map at $\Delta t = \SI{400}{\pico\second}$. (b) Peak energies of the three bands indicated in (a) as a function of $\Delta t$.}
    \label{fig:ExtendedDelayData}
\end{figure}

Time-resolved ARPES results covering an extended pump-probe delay range up to $\Delta t = \SI{400}{\pico\second}$ are compiled in Fig.~\ref{fig:ExtendedDelayData}. Figure~\ref{fig:ExtendedDelayData}(a) shows a ground state ARPES intensity map recorded \SI{10}{\pico\second} before the optical excitation in comparison to a difference intensity map at $\Delta t = \SI{400}{\pico\second}$. The presence of red and blue areas in the difference intensity map indicates that at this time delay the system has still not reached its initial equilibrium state. To further illustrate the long-lived spectral changes, we analyzed the transient peak shifts of three bands labeled (I), (II), and (III) in Fig.~\ref{fig:ExtendedDelayData}(a). The peak positions are determined from Gaussian fits to the transient EDCs of the respective bands. The results of the analysis are summarized in Fig.~\ref{fig:ExtendedDelayData}(b). All bands exhibit rapid energy shifts between \SI{7}{\milli\electronvolt} and \SI{11}{\milli\electronvolt} at time zero followed by an exponential relaxation on a few picosecond time scale. We assign these dynamics to band renormalization processes in response to the excitation and subsequent relaxation of hot carriers. At $\Delta t \approx \SI{5}{\pico\second}$, the energy of band (II) has relaxed back to nearly its initial value, while bands (I) and (III) still exhibit clear energy offsets of $\approx \SI{4}{\milli\electronvolt}$. At the maximum time delay of \SI{400}{\pico\second}, the two bands still show offsets of $\approx \SI{2}{\milli\electronvolt}$. Other time-resolved studies reported on similar long-lived spectral changes in \textit{Td}-\ce{WTe2}~\cite{Sie2019,Dai2015}. Their persistence was associated with heat diffusion out of the excitation volume on a characteristic time scale of several nanoseconds~\cite{Dai2015}.

\section*{Supplementary Note 2: Time resolution and sampling frequencies}

\begin{figure}
    \centering
\includegraphics[width=0.9\linewidth]{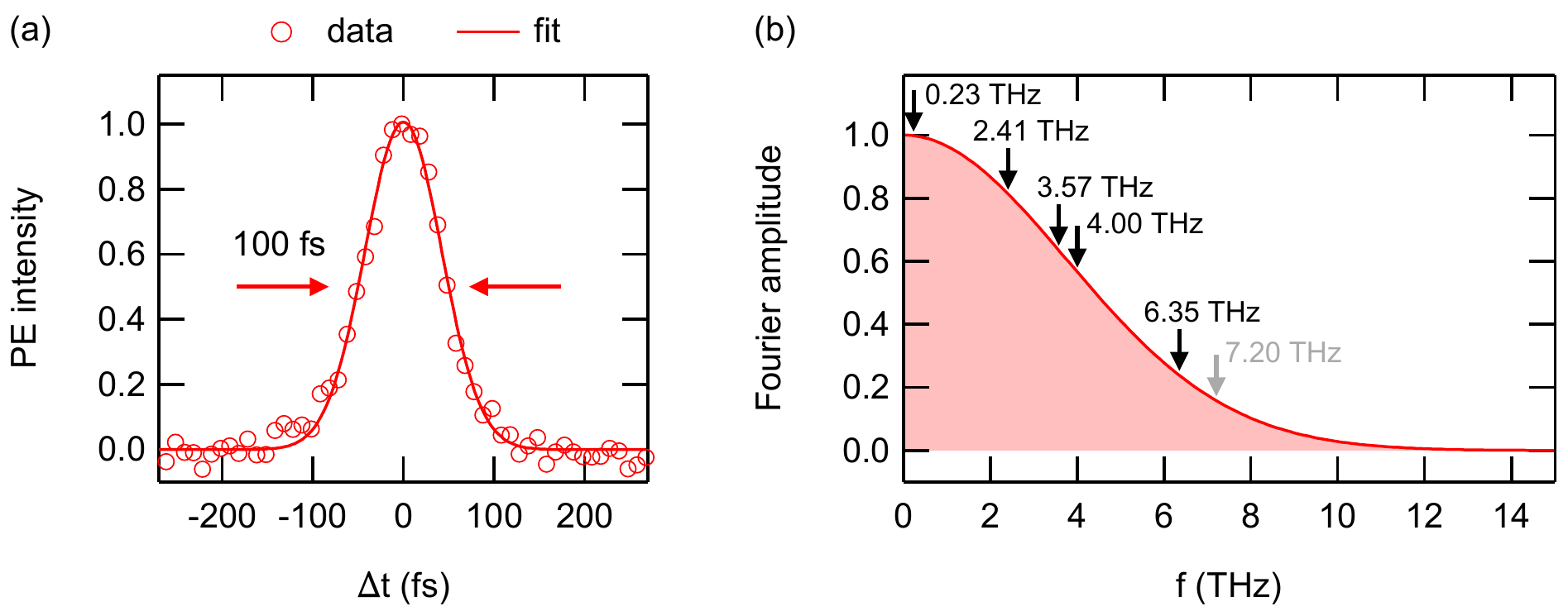}
    \caption{Time resolution of the TRARPES experiment. (a) Experimental NIR-pump/NUV-probe cross-correlation trace measured via photoemission from a polycrystalline gold film at $E-E_F=\SI{1.3}{\electronvolt}$. The solid line is a Gaussian fit to the experimental data. (b) Fourier spectrum of the Gaussian fit in (a). The black markers indicate the coherent phonon frequencies observed in the TRARPES data of \textit{Td}-\ce{WTe2}. The grey marker indicates the highest frequency optical phonon mode of \textit{Td}-\ce{WTe2}.}
    \label{fig:TempRes}
\end{figure}

The coherent phonon frequency range accessible in our TRARPES experiments depends on two factors: the time resolution of the experiment and the sampling rate used for the data acquisition. We determined the time resolution from a pump-probe cross-correlation measurement of a polycrystalline gold film. The cross-correlation trace is shown in Fig.~\ref{fig:TempRes}(a).  It was extracted from TRARPES data for an electron binding energy of $E-E_F = \SI{1.3}{\electronvolt}$ in order to keep the broadening due to finite intermediate state lifetimes small \cite{Bauer2015}. A Gaussian fit to the trace yields a FWHM of $\approx \SI{100}{\femto\second}$ for the time resolution of our setup. \\
In our TRARPES experiments, we measure in the time domain the \underline{convolution} of the sample response (i.e., the coherent phonon-induced oscillations) and the cross-correlation trace representing the finite time resolution. In Fourier space, i.e., in the frequency domain, this corresponds to the \underline{product} of the Fourier spectrum of the sample response and the Fourier spectrum of the cross-correlation trace. Hence, the latter can be considered as a spectral transmission function determining the amplitude at which a phonon peak is probed in the experiment. The Fourier spectrum of the Gaussian fit to the experimental cross-correlation trace is shown in Fig.~\ref{fig:TempRes}(b). Markers indicate the frequencies of the five coherent phonon modes observed in our measurements. Due to the frequency dependence of the transmission function, the detected signal amplitude of the highest frequency mode at \SI{6.35}{\tera\hertz} becomes weakened by a factor of about $4$ in comparison to the lowest frequency mode at \SI{0.23}{\tera\hertz}. However, all five modes are probed at a reasonable amplitude. Even more, also the highest energy optical phonon mode of \textit{Td}-\ce{WTe2} with a frequency of approximately \SI{7.2}{\tera\hertz}~\cite{Kong2015,Jiang2016} would be covered reasonably well by the spectral transmission function. In summary, the available time resolution of the experiment in principle allows for the real-time detection of optical phonons within the entire phonon spectrum supported by \textit{Td}-\ce{WTe2}. For quantitative conclusions on the excitation amplitudes of different phonon modes one should, however, keep in mind that the sensitivity of the experiment decreases with increasing frequency. \\

Independently of the transmission function resulting from the finite time resolution, the accessible frequency range of the experiment also depends on the sampling rate used for the data acquisition and the overall delay range of the measurement. Based on the Nyquist sampling criterion, the sampling frequency must be greater than twice the maximum frequency contained in the signal to be able to identify all frequency components correctly via Fourier transformation. Considering the highest energy phonon branch in \textit{Td}-\ce{WTe2} with a frequency of approximately \SI{7.2}{\tera\hertz}~\cite{Kong2015,Jiang2016}, we therefore chose a minimum sampling rate of \SI{15}{\tera\hertz} for our measurements. \\
For a \underline{fixed} number of data points, increasing the sampling frequency improves the accuracy at which the time-domain signal is traced. At the same time, it worsens the frequency resolution of the Fourier transformation that is given by the reciprocal total sampling time, i.e., the overall delay range chosen in the experiment. To be able to track the high- as well as the low-frequency coherent phonon-induced oscillations \underline{in the time domain}, we recorded TRARPES data sets using two different sampling rates: A sampling rate of \SI{60}{\tera\hertz} enabled us to measure the beating signal caused by the superposition of coherent phonon modes with frequencies between \SI{2.41}{\tera\hertz} ($T = \SI{415}{\femto\second}$) and \SI{6.35}{\tera\hertz} ($T = \SI{157}{\femto\second}$)  [cf. Fig.~2(c) in the main text]. For these measurements, the total delay range was restricted to \SI{3}{\pico\second} to keep the overall data acquisition time at an acceptable level. For the low-frequency interlayer shear mode ($T = \SI{4.35}{\pico\second}$), we chose a significantly longer delay range of \SI{20}{\pico\second} using a sampling rate of \SI{15}{\tera\hertz} to still comply with the Nyquist sampling criterion [cf. Fig.~2(d) in the main text]. Note that similar to conventional frequency domain spectroscopy, it is allowed and (even more) convenient to choose delay steps for accurately probing the time-domain signal that are significantly smaller than the time resolution - in the present study delay steps of \SI{17}{\femto\second} and \SI{67}{\femto\second} at a time resolution of \SI{100}{\femto\second}, respectively.\\

\begin{figure}
    \centering
\includegraphics[width=0.9\linewidth]{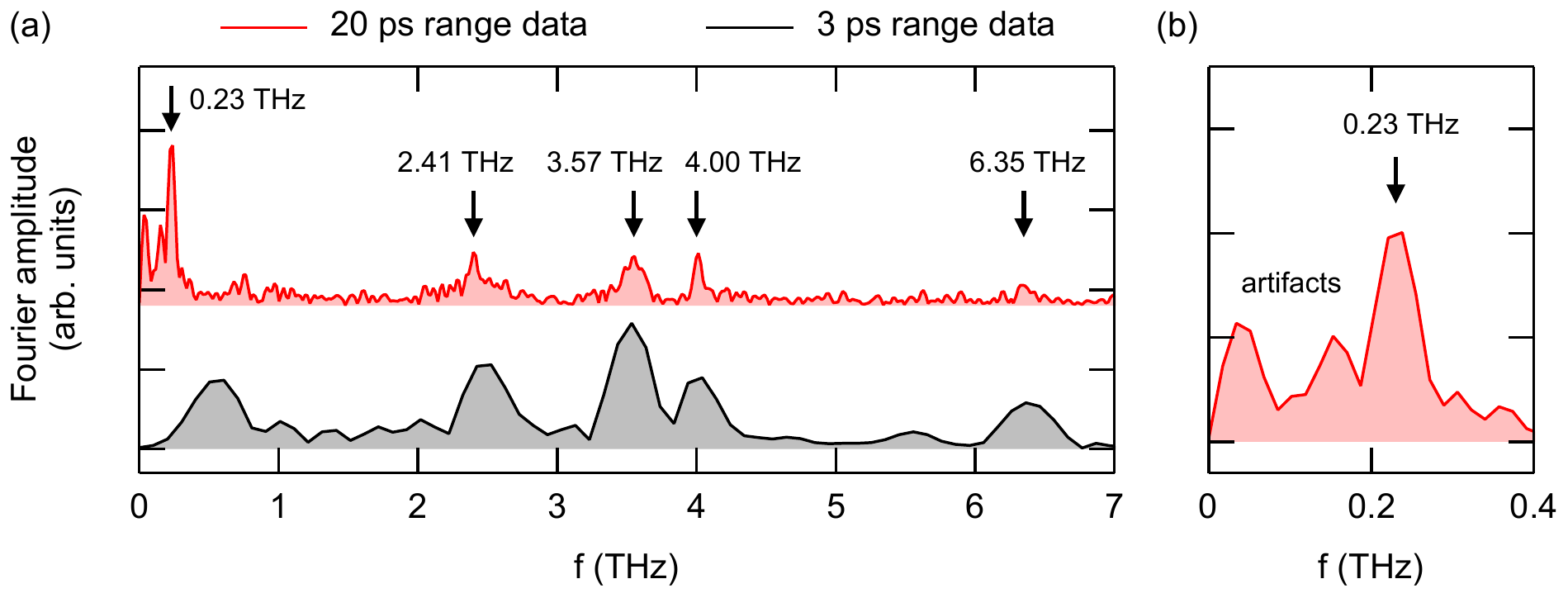}
    \caption{Fourier amplitude spectra of the PE intensity transients shown in Figs.~2(c) and ~2(d) of the main text. (a) Comparison of the Fourier amplitude spectra of the \SI{3}{\pico\second} range data and the \SI{20}{\pico\second} range data. (b) Close-up of the low-frequency range of the \SI{20}{\pico\second} range data.}
    \label{fig:FFTcomparison}
\end{figure}

Figure~\ref{fig:FFTcomparison}(a) compares the Fourier amplitude spectra of the \SI{3}{\pico\second} and \SI{20}{\pico\second} range intensity transients shown in Figs.~2(c) and 2(d) of the main text. The frequency increment of the Fourier analysis, given by the reciprocal total sampling time, is $\approx \SI{0.3}{\tera\hertz}$ and $\approx \SI{0.05}{\tera\hertz}$ for the \SI{3}{\pico\second} and \SI{20}{\pico\second} range data, respectively. While the four high-frequency components are already evident in the amplitude spectrum of the \SI{3}{\pico\second} range data, their shape becomes much clearer in the amplitude spectrum of the \SI{20}{\pico\second} range data allowing for a more precise determination of the peak positions as expected from the discussion above for an extended overall delay range. The \SI{0.23}{\tera\hertz} interlayer shear mode can only be resolved in the \SI{20}{\pico\second} range data set. Furthermore, we observe additional peaks at frequencies below \SI{0.23}{\tera\hertz} [see Fig.~\ref{fig:FFTcomparison}(b)]. They do not match any of the predicted or observed phonon modes of \textit{Td}-\ce{WTe2}~\cite{Jiang2016,Kong2015,Song2016,Dai2015,He2016}. In contrast to the five main frequency peaks between \SI{0.23}{\tera\hertz} and \SI{6.35}{\tera\hertz}, the positions and relative heights of these very-low frequency peaks are quite sensitive to the data analysis procedures, e.g., the delay range included in the analysis and the model function used for background subtraction. We therefore conclude that the appearance of these peaks arises from artifacts of the data analysis.

\section*{Supplementary Note 3: Background correction of Photoemission transients}

The temporal evolution of the transient PE signal of \textit{Td}-\ce{WTe2} is mainly determined by contributions from the hot carrier dynamics and the coherent phonon oscillations. To separate the two signal contributions, we fitted the PE intensity transients with a model function accounting for the carrier population dynamics. Subtraction of the fitting result from the experimental data enables us to extract the pure oscillatory part of the signal. For the fits, we used the following function:
\begin{equation}
I(\Delta t)=\left\{%
\begin{array}{cc}
I_{01}, & \Delta t<t_0 \\
I_{02} + A_1\cdot\exp{(-\frac{\Delta t-t_0}{\tau_1})} + A_2\cdot\exp{(-\frac{\Delta t-t_0}{\tau_2})}, & \Delta t\geq t_0\\
\end{array}%
\right.
\label{eq:fitfunction}
\end{equation}
Here, $I_{01}$ accounts for the constant PE intensity before photoexcitation ($\Delta t<t_0$) and $I_{02}$ accounts for an intensity offset at the maximum investigated delay caused by long-lived spectral changes. At the time $t_0$, excited carrier population and depopulation dynamics set in. They are modeled using a sum of two exponential functions describing an exponential rise ($A_1<0$) and decay ($A_2>0$) of the transient PE intensity with time constants $\tau_1$ and $\tau_2$, respectively. To account for the time resolution $\Delta t_{\text{FWHM}}$ of the experiment, the fitting function was convolved with a Gaussian function $g(t)$:
\begin{equation}
g(t) = \exp{\left(-\frac{4\cdot\ln 2\cdot t^2}{{\Delta {t_\text{FWHM}}}^2}\right)} \text{.}
\end{equation}

Analog fit functions were used to account for the hot carrier response in the peak energy  $E_{\text{band}}(\Delta t)$ and peak width $\Delta E(\Delta t)$ transients.

\begin{figure}
    \centering
\includegraphics[width=0.8\linewidth]{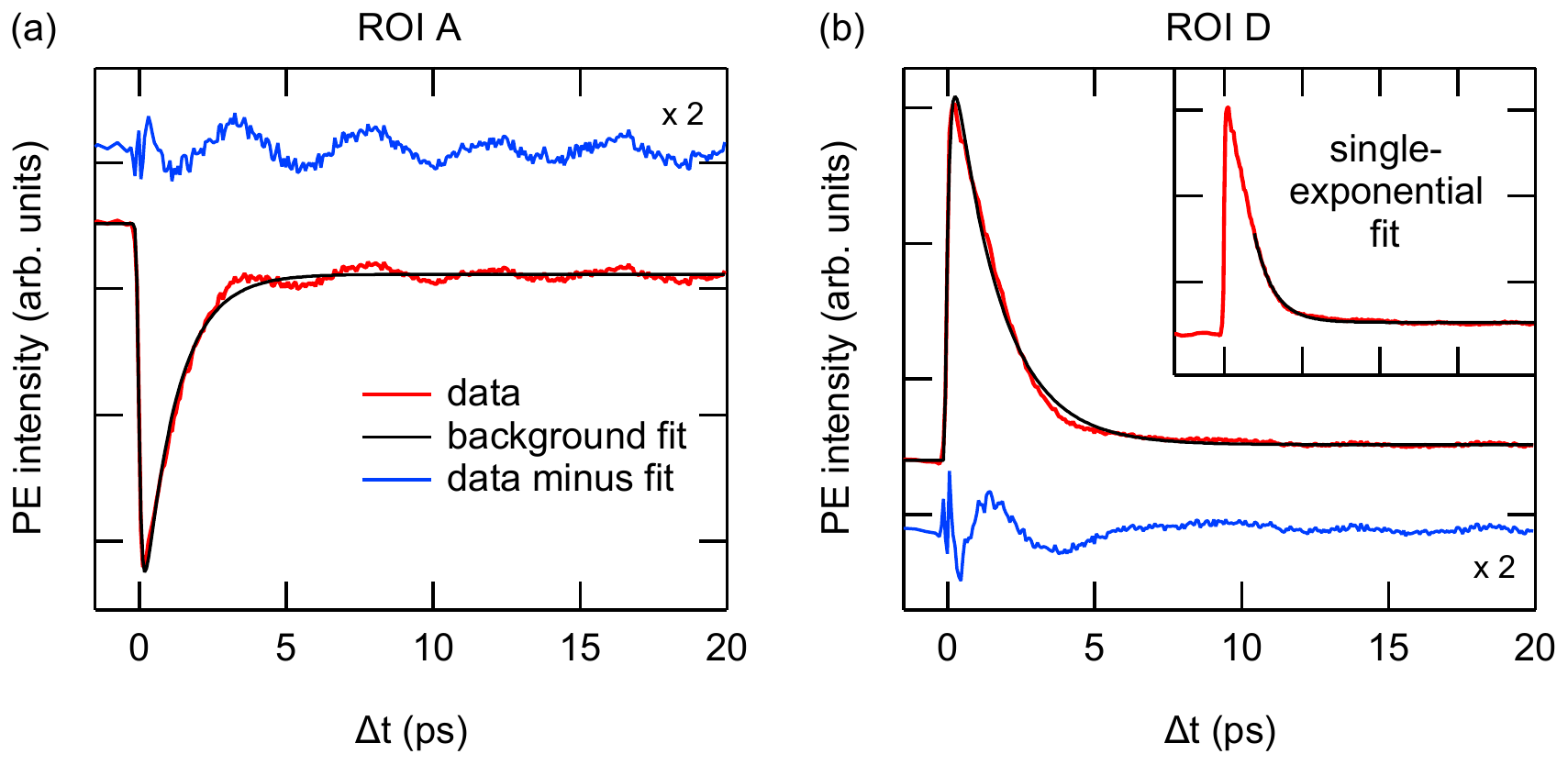}
    \caption{Subtraction of hot carrier contribution. (a) PE intensity transient from ROI A (red) including a fit to the data starting at $\Delta t = \SI{-1.5}{\pico\second}$ according to Eq.~\ref{eq:fitfunction} (black line). The blue line shows the PE intensity transient after subtraction of the fitting result. (b) Same as (a) for ROI D. The inset shows the same PE intensity transient including a single exponential fit starting at $\Delta t = \SI{2}{\pico\second}$.}
    \label{fig:ModelFunction}
\end{figure}

For most cases, Eq.~\ref{eq:fitfunction} is appropriate to account for the dynamics associated with the hot carrier processes. As an example, Fig.~\ref{fig:ModelFunction}(a) shows the PE intensity transient for ROI A in Fig.~4(c) of the main text and the corresponding fit using equation~\ref{eq:fitfunction}. However, in regions near $E_\text{F}$, we observed deviations from a pure exponential signal decay, including the Weyl point region [ROI D, cf. Fig.~4(d) of the main text]. The PE intensity transient from this region is shown in Fig.~\ref{fig:ModelFunction}(b) including the result of a fit using equation~\ref{eq:fitfunction}. The deviation from an exponential decay is clearly visible and results in a signal distortion at small delays in the hot carrier-corrected data shown as blue line. The quality of the background subtraction procedure can be improved for this type of data by restricting the delay range to $\Delta t > \SI{2}{\pico\second}$ and by describing the hot carrier processes with a single exponential fit function. Still, a complete removal of hot carrier contributions is not possible resulting in a residual signal visible as a distortion in Fig.~4(d) between $\Delta t = \SI{5}{\pico\second}$ and $\Delta t = \SI{10}{\pico\second}$.

\section*{Supplementary Note 4: Compilation of Fourier maps}

For the energy- and momentum-resolved Fourier analysis of the data, the ARPES intensity maps were divided into small integration regions with a size of $\SI{23}{\milli\electronvolt} \times \SI{0.01}{\per\angstrom}$ [Figs. 3(a) and 4(c)] or $\SI{6}{\milli\electronvolt} \times \SI{0.0025}{\per\angstrom}$ [Fig. 5(a)]. Each of these regions features a PE intensity transient showing a superposition of carrier population dynamics and coherent phonon-induced oscillations. Some spectral regions show an additional signal confined to the delay range around time zero [cf. Fig.~4(d) and 4(e) of the main text] that we assign to the pump-probe laser cross-correlation. We therefore restricted the data range for the analysis to a time delay range starting well after this signal contribution. Before the Fourier transformation, the PE intensity transients from each region were fitted for background subtraction with a single-exponential function to account for the carrier relaxation dynamics. The \SI{3}{\pico\second} range data were analyzed for a delay range starting at $\Delta t = \SI{280}{\femto\second}$. In the case of the \SI{20}{\pico\second} range data, we had to exclude the first \SI{2}{\pico\second} after time zero as a single-exponential fit of the intensity transients especially around the Fermi level results otherwise in low-frequency artifacts that interfere with the \SI{0.23}{\tera\hertz} oscillation [see intensity transient of ROI D shown in Fig.~\ref{fig:ModelFunction}(b)].

\begin{figure}
    \centering
\includegraphics[width=0.9\linewidth]{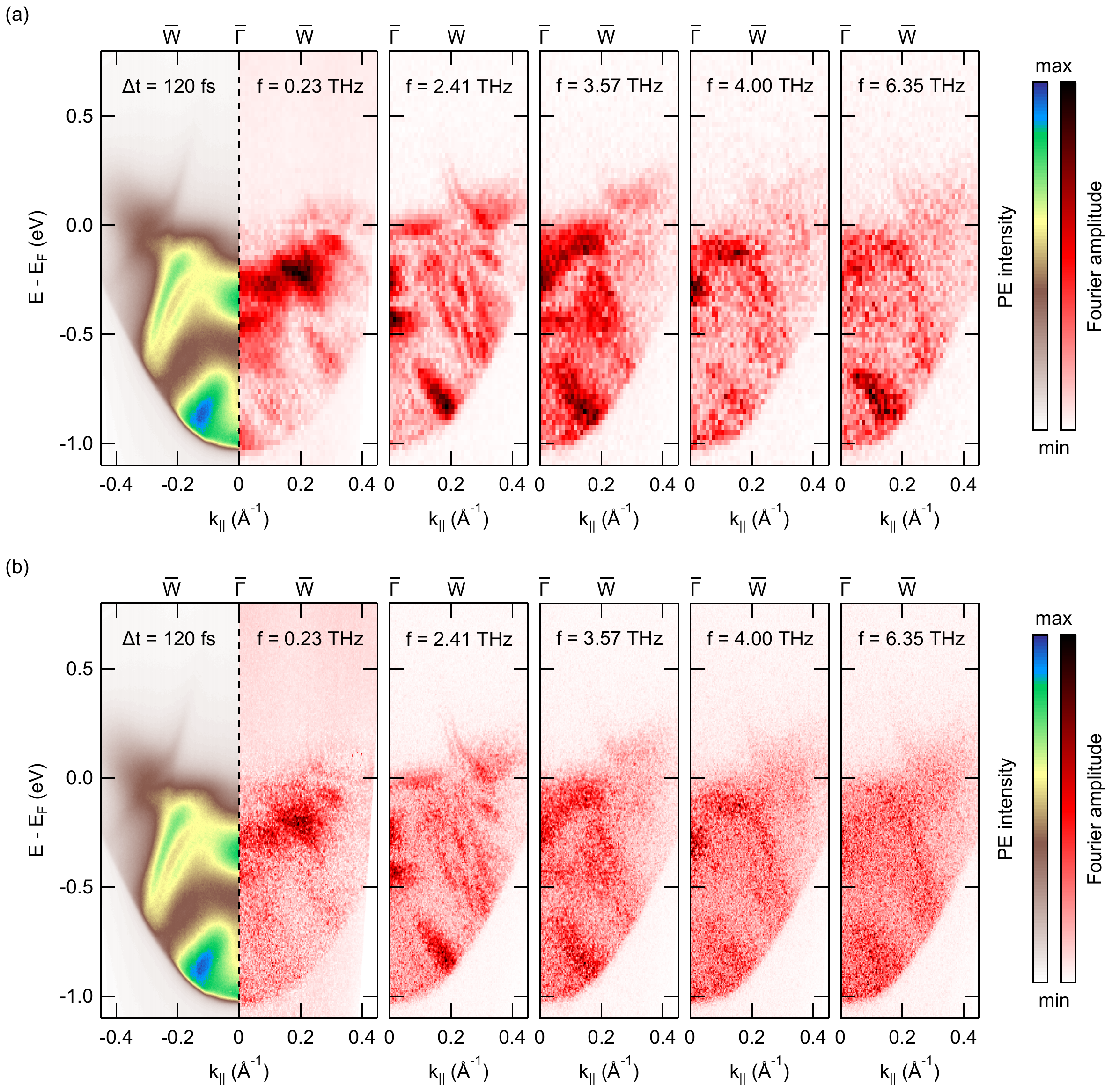}
    \caption{Phonon mode-selective Fourier amplitude analysis of the TRARPES data. ARPES intensity map at $\Delta t = \SI{120}{\femto\second}$ in comparison to Fourier maps of the five phonon frequencies identified in the Fourier amplitude spectra. (a) Fourier maps generated from ARPES intensity maps that were binned over regions of $\SI{23}{\milli\electronvolt} \times \SI{0.01}{\per\angstrom}$. (b) Fourier maps generated from ARPES intensity maps that were binned over regions of $\SI{6}{\milli\electronvolt} \times \SI{0.0025}{\per\angstrom}$.}
    \label{fig:FFTMaps}
\end{figure}

After subtraction of the single-exponential background, the resulting transients were zero-padded to triple their lengths before their amplitude spectra were determined via fast Fourier transformation. This results in an energy- and momentum-dependent data set of Fourier amplitude spectra. A cut through this data set at a fixed frequency represents the energy- and momentum-resolved Fourier amplitude at the selected frequency. It can be displayed as a color-coded image which we refer to as Fourier map. Phonon-mode selectivity is achieved by selecting the frequencies corresponding to the coherent phonon modes. As an extended version of Fig.~3 in the main text, Fig.~\ref{fig:FFTMaps} shows Fourier maps at \SI{0.23}{\tera\hertz}, \SI{2.41}{\tera\hertz}, \SI{3.57}{\tera\hertz}, \SI{4.00}{\tera\hertz}, and \SI{6.35}{\tera\hertz} in comparison to the excited state ARPES intensity map at $\Delta t = \SI{120}{\femto\second}$. The \SI{0.23}{\tera\hertz} Fourier map results from the analysis of the \SI{20}{\pico\second} range data. In contrast, the four high-frequency Fourier maps are derived from the \SI{3}{\pico\second} range data.

We finally would like to add that the details of the Fourier map patterns critically depend on the type of spectral changes in the ARPES data that are induced by the coherent phonon excitation. In general, a pure periodic modulation in the PE intensity results in a slowly varying Fourier amplitude in the energy and momentum range of the signal modulations. In contrast, changes in peak energy or peak width result in distinct minima in the Fourier amplitude right at the peak position and Fourier amplitude maxima on both sides next to the peak. These effects are best seen when comparing the \SI{0.23}{\tera\hertz} and \SI{2.41}{\tera\hertz} Fourier maps. As discussed in the main text, the oscillations in the \SI{0.23}{\tera\hertz} data predominantly arise from PE intensity modulations resulting in a rather smooth Fourier map pattern. In contrast, the \SI{2.41}{\tera\hertz} Fourier amplitude shows a distinct fine structure of amplitude mimima that are indicative for shifts in the band energies or changes in the band widths in response to the excitation of the \SI{2.41}{\tera\hertz} phonon mode.

\section*{Supplementary Note 5: Resolution enhancement}

\begin{figure}
    \centering
\includegraphics[width=0.8\linewidth]{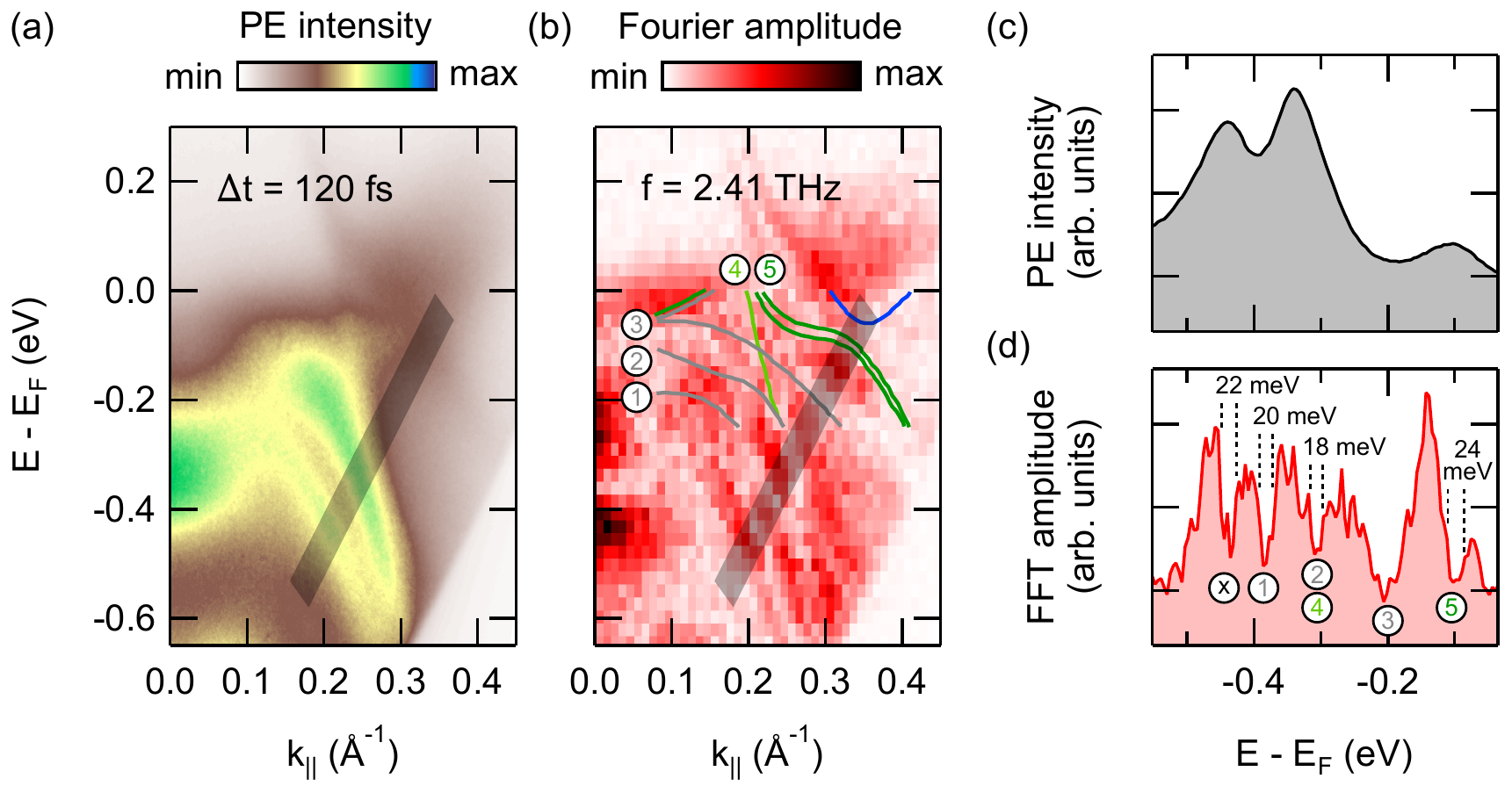}
    \caption{Resolution enhancement in the Fourier maps. Close-ups of  (a) the ARPES intensity map and (b) the \SI{2.41}{\tera\hertz} Fourier map shown in Fig. 3(a) of the main text with the energy-momentum cut analyzed in (c) and (d) indicated. The Fourier map was generated from ARPES intensity maps that were binned over regions of $\SI{23}{\milli\electronvolt} \times \SI{0.01}{\per\angstrom}$. It is overlaid with experimental band dispersions from Ref. \cite{Bruno2016}. Surface bands were omitted for the sake of clarity. (c) Energy distribution curve extracted from the ARPES intensity map along the energy-momentum cut indicated in (a). (d) Fourier amplitude as a function of binding energy along the energy-momentum cut indicated in (b). The data were extracted from a Fourier map generated from ARPES data that were binned over regions of $\SI{6}{\milli\electronvolt} \times \SI{0.0025}{\per\angstrom}$. The numbers indicate the assignment of the Fourier amplitude dips to the different bands shown in (b). 'x' indicates a band signature visible in the Fourier map that is outside the energy range that was probed in Ref. \cite{Bruno2016}. Spectral widths of selected dips are indicated.} 
    \label{fig:ResEnhancement}
\end{figure}

The effect of resolution enhancement in the Fourier maps is illustrated in Fig. \ref{fig:ResEnhancement}. Figure \ref{fig:ResEnhancement}(a) shows a close-up of the ARPES intensity map in Fig. 3(a) of the main text with a selected energy-momentum cut indicated by the grey-shaded bar. The EDC along this cut [Fig. \ref{fig:ResEnhancement}(c)] exhibits three peaks with widths in the order of \SI{100}{\milli\electronvolt} governed by the experimental resolution of \SI{40}{\milli\electronvolt} and considerable thermal broadening due to the sample temperature of \SI{300}{\kelvin}. Figure \ref{fig:ResEnhancement}(b) shows for comparison the corresponding \SI{2.41}{\tera\hertz} Fourier map exhibiting a distinct fine structure not visible in the ARPES intensity map. The data is overlaid with experimental band dispersions from a high-resolution ARPES study conducted with an energy resolution of \SI{2}{\milli\electronvolt} and at a sample temperature below \SI{20}{\kelvin} \cite{Bruno2016}. In the overlapping energy-momentum regime, the experimental band dispersions exhibit a striking agreement with the dispersion of the fine structure. Even more, an intuitive extrapolation of part of the data from Ref. \cite{Bruno2016} to higher binding energies also seems to follow the fine structure in the \SI{2.41}{\tera\hertz} Fourier map. Fig. \ref{fig:ResEnhancement}(d) displays the Fourier amplitude as a function of binding energy along the selected energy-momentum cut, in which the fine structures appears as distinct and narrow amplitude dips. The spectral width of these dips is typically in the order of \SI{20}{\milli\electronvolt} and, therefore, much smaller than the width of the characteristic spectral signatures in the ARPES intensity maps and also smaller than the spectral resolution of our experiment. Along the selected energy-momentum cut, we identify in the Fourier amplitude signal five dips in comparison to three peaks in the EDC, implying that the obvious resolution enhancement allows observing details that are hidden in the ARPES data due to strong spectral broadening. The assignment of the dips to different bands is indicated by numbers [compare Fig.~\ref{fig:ResEnhancement}(b)]. Bands number 1, 2, and 4 were assigned based on an extrapolation of the high resolution ARPES data to higher binding energies, while the assignment of bands number 3 and 5 results from the direct comparison of the Fourier map with the band dispersions. The amplitude dip  labeled 'x' at $E-E_F \approx \SI{-0.35}{\milli\electronvolt}$ could not be assigned to any of the indicated bands as the related fine structure is completely outside the energy range probed in Ref. \cite{Bruno2016}. We assume, however, that also this fine structure follows the dispersion of an electronic band of \textit{Td}-\ce{WTe2}.

\begin{figure}
    \centering
    \includegraphics[width=0.75\linewidth]{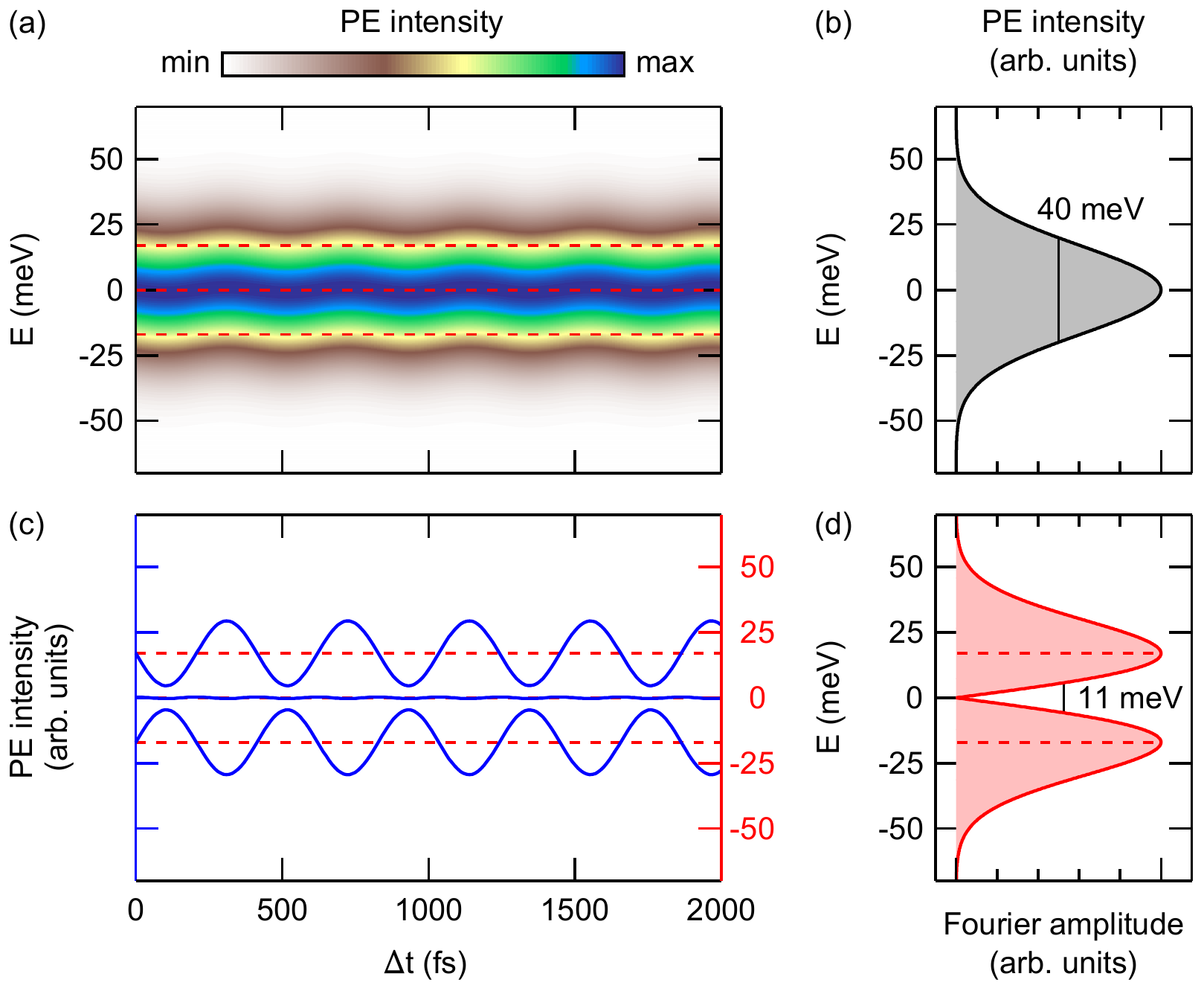}
    \caption{Origin of resolution enhancement. (a) Simulated PE intensity transient of the Gaussian peak shown in (b). The peak energy is periodically modulated with an amplitude of \SI{1}{\milli\electronvolt} and a frequency of \SI{2.41}{\tera\hertz}. (c) Changes in the PE intensity as a function of $\Delta t$ (blue lines) at three selected energies indicated by the dashed red lines. The dashed red lines also indicate the respective zero of the changes in the PE intensity. (d) \SI{2.41}{\tera\hertz} Fourier amplitude as a function of energy determined from a Fourier transformation of the PE intensity transient shown in (a).}
    \label{fig:ResEnhance_Bandshift}
\end{figure}

The origin of the resolution enhancement in the Fourier maps is schematically illustrated by a simulation presented in Fig. \ref{fig:ResEnhance_Bandshift}. We considered a Gaussian peak with a width of \SI{40}{\milli\electronvolt} (FWHM) [Fig. \ref{fig:ResEnhance_Bandshift}(b)] that is subject to a periodic peak energy shift with an amplitude of \SI{1}{\milli\electronvolt} and a frequency of \SI{2.41}{\tera\hertz} [Fig. \ref{fig:ResEnhance_Bandshift}(a)]. A quantitative analysis of the oscillation amplitude shows that near the peak maximum the changes in the PE intensity are considerably smaller than in large parts of the flanks of the Gaussian peak [Fig. \ref{fig:ResEnhance_Bandshift}(c)]. This energy dependence directly translates into the \SI{2.41}{\tera\hertz} Fourier amplitude signal, which is shown in Fig. \ref{fig:ResEnhance_Bandshift}(d). At the energy of the peak maximum, the Fourier amplitude exhibits a distinct dip with a width (FWHM) that is significantly smaller than the width of the oscillating PE signal. In the present example, we observe in the Fourier signal in comparison to the PE signal a reduction of the width by a factor of $\approx 4$ to a value of \SI{11}{\milli\electronvolt}. Note, however, that the situation is in general more complex as different dynamical processes such as peak shift, peak broadening, and an overall modulation of spectral weight can independently superimpose each other.

\section*{Supplementary Note 6: Time-dependent Fourier amplitude and phase analysis}

A time-dependent Fourier amplitude analysis enables us to gain information on the lifetime of the different coherent phonons and to identify possible mode frequency shifts as a function of time. Furthermore, a time-dependent Fourier phase analysis of phonon-induced band shifts allows for conclusions on the underlying excitation mechanisms of the coherent phonons, i.e., the distinction between a displacive excitation (DECP) scenario~\cite{Zeiger1992} and an impulsive stimulated Raman scattering (ISRS) excitation scenario~\cite{Garrett1996}.

\begin{figure}
    \centering
\includegraphics[width=1\linewidth]{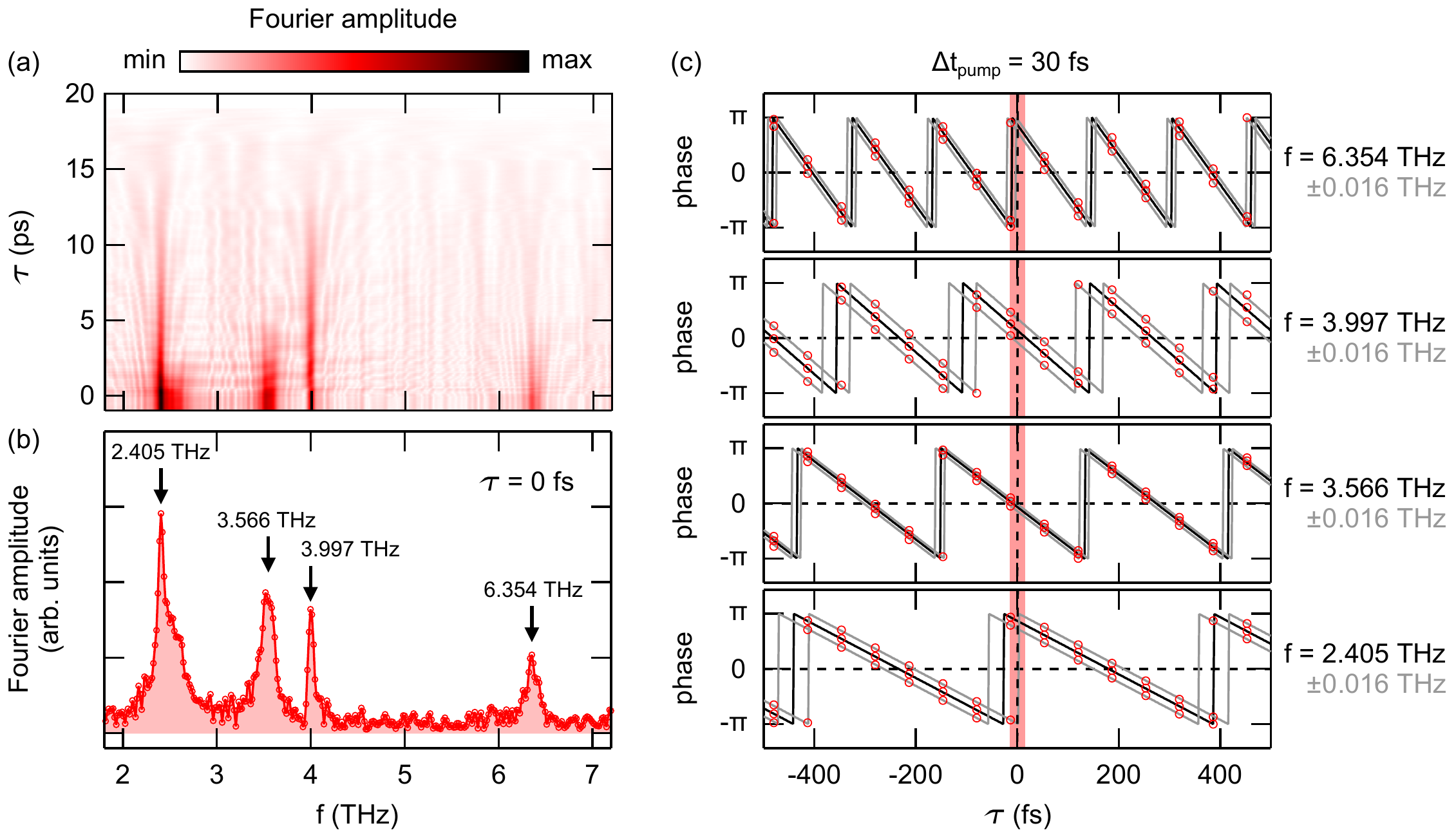}
    \caption{Time-resolved Fourier amplitude and phase analysis of the high-frequency modes. (a) Fourier amplitude as a function of frequency and starting point $\tau$ of the band energy transient shown in Fig.~3(b) of the main text. Details on the evaluation of the data are given in the text. (b) Fourier amplitude spectrum extracted from (a) for $\tau = \SI{0}{\femto\second}$. (c) Fourier phase as a function of $\tau$ for the peak positions of the four high-frequency modes. The black and gray lines show fits of a sawtooth function to the peak maximum data and the data of neighboring frequencies, respectively. The red shaded area at $\tau = \SI{0}{\femto\second}$ indicates the uncertainty in determining the time zero of the experiment due to the temporal width of the pump pulse.}
    \label{fig:Phasenauswertung}
\end{figure}

Figure~\ref{fig:Phasenauswertung} shows results of a time-dependent amplitude and phase analysis of the band energy transient of the prominent feature centered at $\overbar{\Gamma}$ and $E - E_\text{F} \approx \SI{-0.3}{\electronvolt}$ in the ARPES intensity maps. The data comprise the frequency range of the four high-frequency coherent phonon modes, which periodically modulate the peak energy as discussed in the main text [cf. Figs.~3(b) and 3(c)]. For a time-dependent analysis, we successively shortened the band energy transient shown in Fig.~3(b) of the main text at its beginning and performed a Fourier transformation of the remaining transient, respectively. Fig.~\ref{fig:Phasenauswertung}(a) displays the resulting Fourier amplitude as a function of frequency and starting point $\tau$ of the analyzed transient. The results illustrate the distinct differences in the lifetimes of the different modes, which obviously also conform with their differences in spectral width. The \SI{3.57}{\tera\hertz} mode and the \SI{6.35}{\tera\hertz} mode become damped within less than \SI{6}{\pico\second}, whereas the \SI{2.41}{\tera\hertz} mode and the \SI{4.00}{\tera\hertz} mode are still visible at \SI{12}{\pico\second}. For the \SI{2.41}{\tera\hertz} mode, we additionally observe a fast decaying shoulder which may hint to the excitation of a neighboring short-living coherent phonon. The data do not show any indications for a spectral shift of the modes as a function of time.

Figure~\ref{fig:Phasenauswertung}(b) shows the Fourier amplitude spectrum for $\tau = \SI{0}{\femto\second}$ extracted from Fig.~\ref{fig:Phasenauswertung}(a). We used this spectrum to determine the positions of the peak maxima (indicated in the figure by the arrows) for which we performed a time-dependent Fourier phase analysis. The results of the phase analysis are shown in Fig.~\ref{fig:Phasenauswertung}(c). We additionally performed a phase analysis for the data points next to the positions of the peak maxima at $\pm \SI{0.016}{\tera\hertz}$ to estimate the error in the determined phase. All results are fitted with sawtooth functions with the amplitudes set to a value of $\pi$. The phases of the oscillations at time zero indicate the underlying phonon excitation mechanism. Time zero was determined from the peak position of the laser cross-correlation signal that is in part of the data visible during the temporal overlap of pump and probe pulse [cf. Fig.~4(d) and 4(e) in the main text]. Our analysis yields initial phases of $(0.87 \pm 0.14)\pi$ for the excitation of the \SI{2.41}{\tera\hertz} mode, $(-0.07 \pm 0.06)\pi$ for the excitation of the \SI{3.57}{\tera\hertz} mode, $(0.14 \pm 0.22)\pi$ for the excitation of the \SI{4.00}{\tera\hertz} mode, and $(0.85 \pm 0.12)\pi$ for the excitation of the \SI{6.35}{\tera\hertz} mode. Within the error bars and under consideration of the pump pulse width, these results are compatible with a cosinusoidal response of the band energies to the excitation of the coherent phonons implying for all cases a displacive excitation~\cite{Zeiger1992}.

For the phase analysis of the low-frequency interlayer shear mode, we analyzed the band energy transient of ROI A$'$ as shown in Fig.~4(e) of the main text. In this case, the initial phase can be directly determined from a fit of the transient. Although overlaid with a laser cross-correlation PE signal around time zero, the oscillation can be clearly identified as sinusoidal [cf. black line in Fig.~4(e) of the main text], implying for this coherent phonon mode an impulsive excitation~\cite{Garrett1996}.

\section*{Supplementary Note 7: Estimation of the shear mode amplitude}

An approximate estimation of the shear mode amplitude in \textit{Td}-\ce{WTe2} resulting from the excitation with \SI{827}{\nano\meter} pump pulses at an incident fluence of \SI{110}{\micro\joule\per\square\centi\meter} becomes possible under consideration of quantitative results on the shear displacement reported for excitation with \SI{23}{\tera\hertz} pulses in Ref.~\cite{Sie2019}. To account for the different excitation wavelengths, we first referenced our data to fluence-dependent data on the shear mode frequency reported in a TRR study of \textit{Td}-\ce{WTe2} using \SI{800}{\nano\meter} pump pulses~\cite{He2016}. Corresponding data for \SI{}{\tera\hertz} excitation as a function of the incident total pulse energy are also found in in Ref.~\cite{Sie2019}. Both studies report on an overall shift of the shear mode frequency of approximately \SI{16}{\percent} over the probed pump fluence and energy ranges, respectively. Comparison of the data sets allow, therefore, to reference the \SI{800}{\nano\meter} fluence values to the \SI{23}{\tera\hertz} pulse energy values. This comparison yields that the excitation with $\SI{800}{\nano\meter}$, $\SI{150}{\micro\joule\per\square\centi\meter}$ pump pulses results in a shear amplitude approximately corresponding to the excitation with \SI{23}{\tera\hertz} pump pulses with a field amplitude of $\SI{0.65}{\mega\volt\per\centi\meter}$. Based on the quantitative discussion on the shear displacement in Ref.~\cite{Sie2019}, this allows to roughly estimate the shear mode amplitude in our experiment to a value in the range of \SI{1}{\pico\meter}.
In this context, it is noteworthy that in Ref.~\cite{Sie2019} a complete annihilation of the Weyl points is predicted already for a shear displacement of \SI{2}{\pico\meter}, even though the complete transition into the centrosymmetric $1\textit{T}'(^*)$ phase requires an overall shear displacement of $\approx \SI{12}{\pico\meter}$.

\section*{Supplementary Note 8: Comparison of the \SI{0.23}{\tera\hertz} data with band dispersions including surface bands}

\begin{figure}
    \centering
\includegraphics[width=0.7\linewidth]{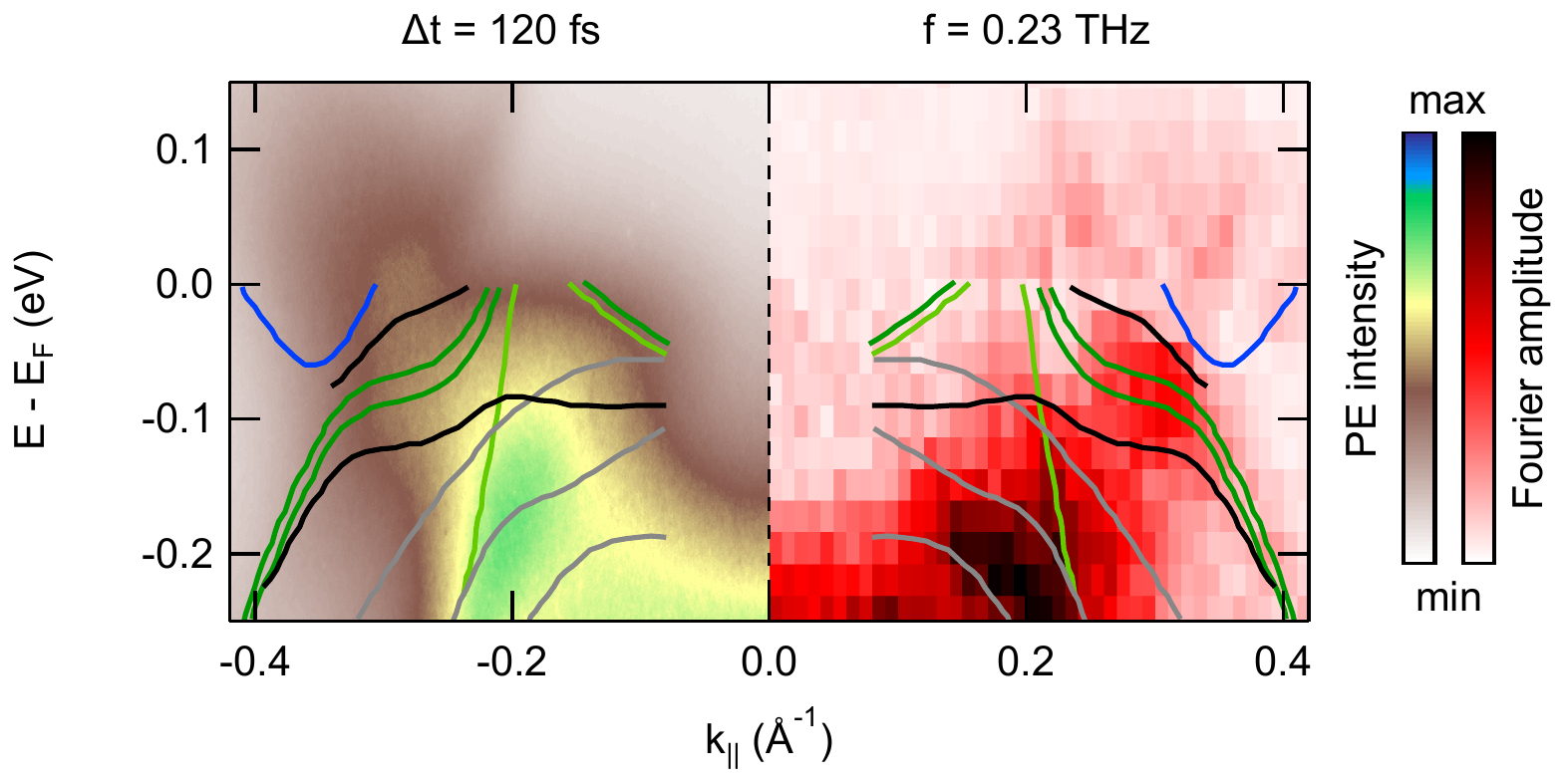}
    \caption{Close-ups of the ARPES intensity map near $E_\text{F}$ at $\Delta t = \SI{120}{\femto\second}$ (left) and the corresponding \SI{0.23}{\tera\hertz} Fourier map (right). The data is overlaid with experimental band dispersions from Ref.~\cite{Bruno2016}. In contrast to Fig. 4(c) of the main text, surface bands are included in black.}
    \label{fig:Surface_States}
\end{figure}

Figure \ref{fig:Surface_States} shows close-ups of an ARPES intensity map at $\Delta t = \SI{120}{fs}$ and the corresponding \SI{0.23}{\tera\hertz} Fourier map overlaid with experimental band dispersions from Ref. \cite{Bruno2016} with the surface bands included. \\

\section*{Supplementary Note 9: Comparison of ROI above $E_\text{F}$}

\begin{figure}
    \centering
\includegraphics[width=0.9\linewidth]{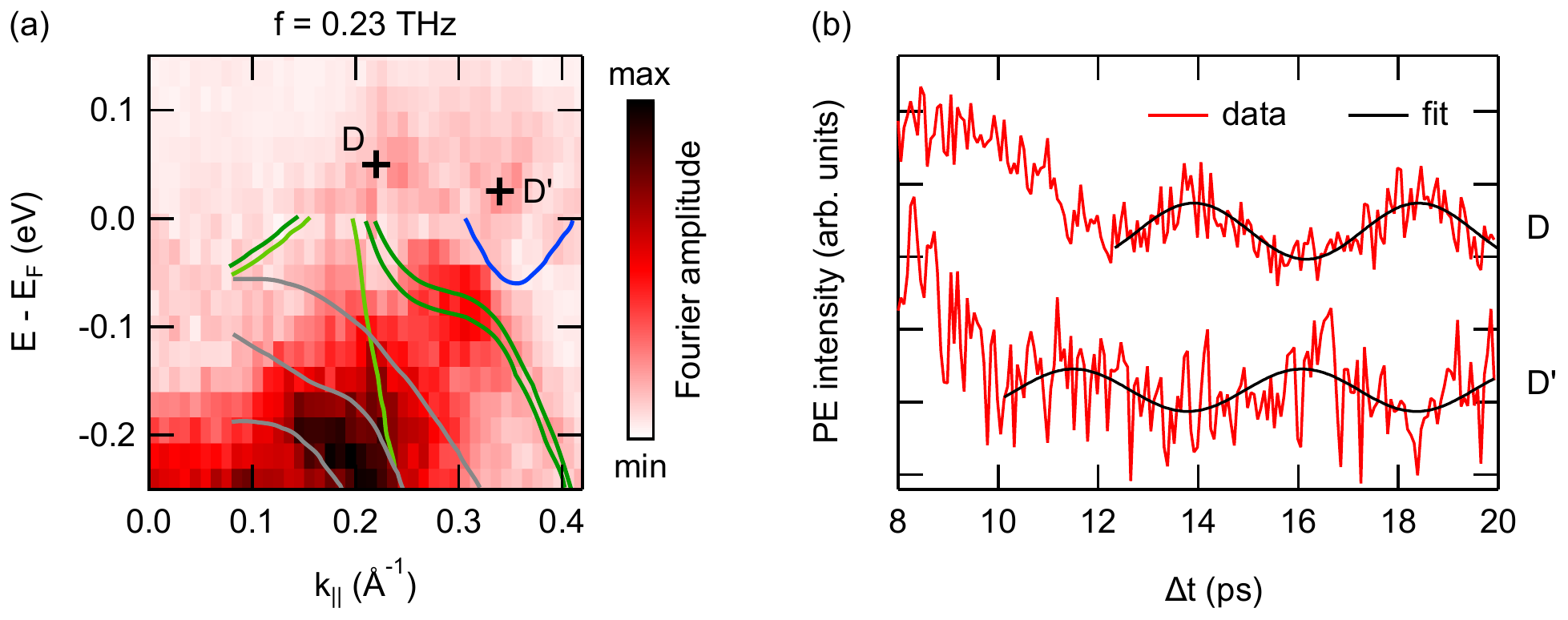}
    \caption{PE intensity transients from energy-momentum areas above $E_\text{F}$. (a) \SI{0.23}{\tera\hertz} Fourier map with ROI D and ROI D$'$ indicated. (b) Comparison of PE intensity transients from ROI D and ROI D$'$. The black lines are fits of a sinusoidal function to the data. For the analysis, we chose a signal integration area of $\SI{50}{\milli\electronvolt} \times \SI{0.025}{\per\angstrom}$.}
    \label{fig:ROI_comparison}
\end{figure}

Figure~\ref{fig:ROI_comparison} compares PE intensity transients from the two energy-momentum areas above $E_\text{F}$ that exhibit distinct amplitude maxima in the \SI{0.23}{\tera\hertz} Fourier map. In Fig.~\ref{fig:ROI_comparison}(a), the two areas (ROI D and ROI D$'$) are indicated in the Fourier map, with ROI D already introduced as the Weyl point area in the main text. The corresponding PE transients are shown in Fig.~\ref{fig:ROI_comparison}(b). Whereas the signal from  ROI D$'$ oscillates in phase with the signal from the ROI below $E_\text{F}$, ROI D shows a characteristic anti-phase behavior as already discussed in the main text.

\section*{Supplementary Note 10: Simulation of the Fourier amplitude modulation observed in ROI C}

\begin{figure}
    \centering
    \includegraphics[width=0.9\linewidth]{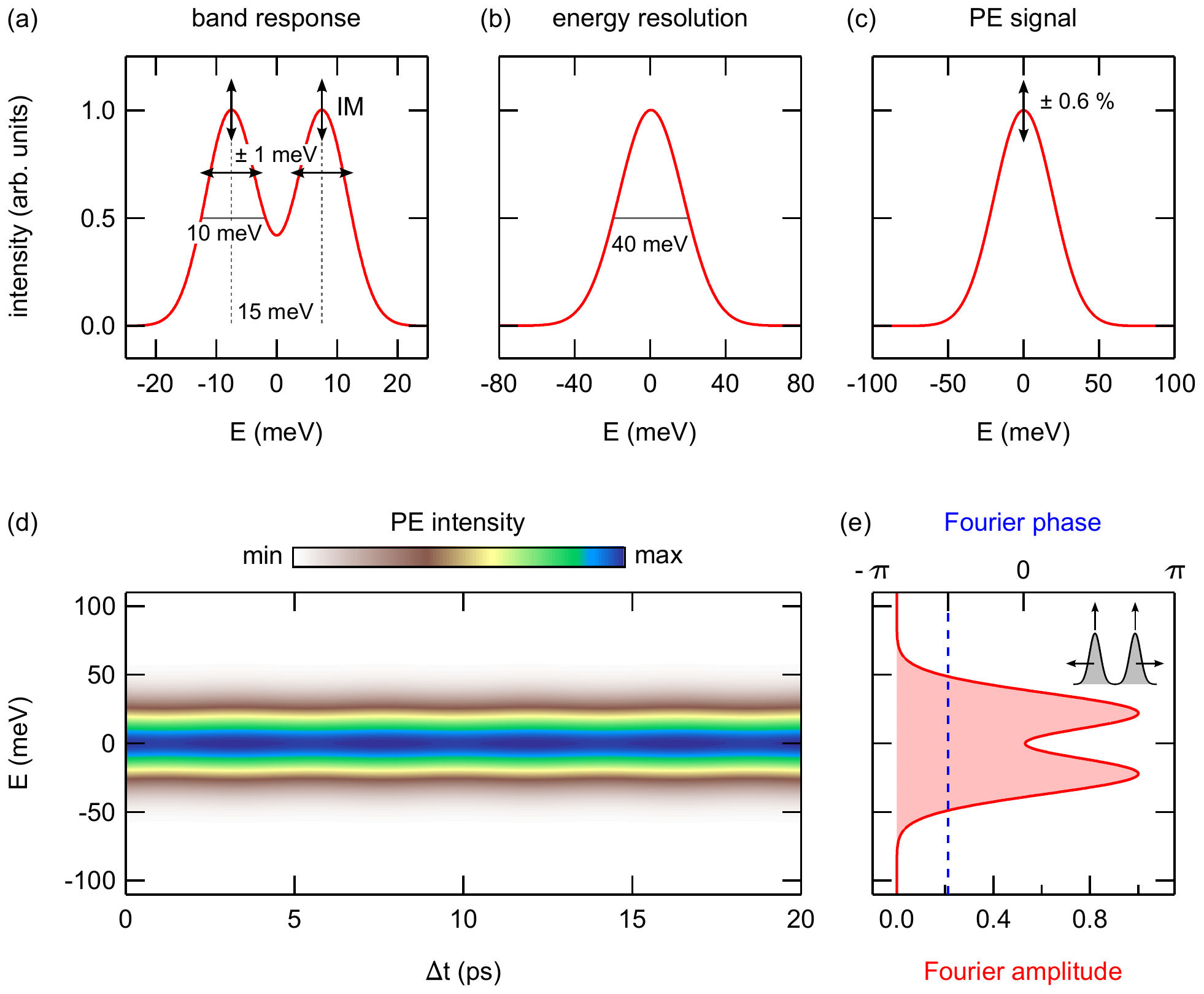}
    \caption{Schematic illustration of the procedure used to simulate the \SI{0.23}{\tera\hertz} Fourier amplitude modulation of the upper hole pocket. The spin splitting of the bands was modelled by two Gaussians (a), which were convolved by a Gaussian representing the energy resolution of the TRARPES experiment (b), yielding the PE signal (c). The parameters chosen for the simulations are indicated. The amplitude of the peak intensity modulation (IM) was adjusted until the resulting PE signal yielded a relative oscillation amplitude of $\pm \SI{0.6}{\percent}$. (d) Simulated PE intensity transient for scenario (ii). The corresponding energy-dependent Fourier amplitude and phase signals (e) at the modulation frequency of \SI{0.23}{\tera\hertz} were determined from a Fourier transformation of the PE intensity transient (d).}
    \label{fig:Simulation}
\end{figure}

The procedure used to simulate the \SI{0.23}{\tera\hertz} Fourier amplitude modulation of the upper hole pocket observed near ROI C including the values chosen for the free parameters is schematically illustrated in Fig. \ref{fig:Simulation}. The spin splitting of the bands was modeled by two Gaussians separated by \SI{15}{\milli\electronvolt} in accordance with results of high resolution ARPES data \cite{Bruno2016}. The peak width (FWHM) was set to \SI{10}{\milli\electronvolt}, simulations with different values showed, however, that this parameter does not critically affect our results. For modeling the dynamical response, we assumed that peak energies and peak intensities are modulated \underline{independently} by the excitation of the shear mode. Peak energy shifts can result from band renormalization effects due to electron-phonon interaction as the crystal structure is periodically modulated by the excitation of a coherent phonon. Changes in the peak intensity (as detected in a photoemission experiment) can arise from changes in transition matrix elements, which are for instance sensitive to the crystal symmetry \cite{Huefner1995}. Oscillation amplitudes for rigid band shift [scenario (i)] and changes in the spin splitting [scenario (ii) and (iii)] were set to \SI{1}{\milli\electronvolt}. This value was estimated from the $\approx\SI{15}{\milli\electronvolt}$ equilibrium spin splitting of the bands \cite{Bruno2016} and under consideration of the approximate shear mode amplitude of \SI{1}{\pico\meter} at the applied excitation density (see Supplementary Note 7) with the latter value corresponding to $\approx\SI{10}{\percent}$ of the shear displacement that transforms the $\textit{Td}$ structure into the centrosymmetric $1\textit{T}'(^*)$ structure \cite{Sie2019}. To account for the energy resolution of the experiment, the band response [Fig. \ref{fig:Simulation}(a)] was convolved with a Gaussian with a width of \SI{40}{\milli\electronvolt} (FWHM) [Fig. \ref{fig:Simulation}(b)] to simulate the PE signal [Fig. \ref{fig:Simulation}(c)]. The oscillation amplitude of the peak intensities (IM) was adjusted until the relative amplitude of the simulated PE intensity oscillation yielded a value of $\pm \SI{0.6}{\percent}$ in agreement with the experimental data from ROI C shown in Fig.~4(d) of the main text.
The simulated PE intensity transient and its corresponding energy-dependent Fourier amplitude and phase signals at the modulation frequency of \SI{0.23}{\tera\hertz} are exemplary shown in Figs. \ref{fig:Simulation}(d) and \ref{fig:Simulation}(e) for scenario (ii).

\section*{Supplementary Note 11: Band dispersion along $\Gamma$-Z}

\begin{figure}
    \centering
\includegraphics[width=0.9\linewidth]{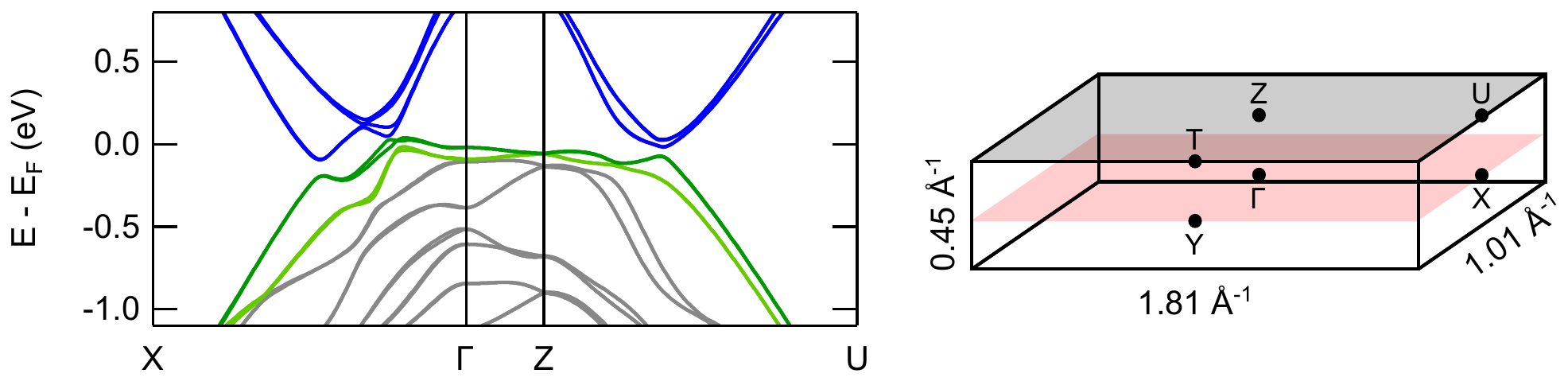}
    \caption{Results of the band structure calculation of \textit{Td}-\ce{WTe2} for the $\Gamma$-X, $\Gamma$-Z, and Z-U direction. The 3D Brillouin zone of \textit{Td}-\ce{WTe2} is shown to the right.}
    \label{fig:berechneteBS}
\end{figure}

In the main text, we discuss the potential relevance of the dispersion of the two low energy bulk bands along the $\Gamma$-Z direction. For comparison, we show in Fig. \ref{fig:berechneteBS} results of the band structure calculation for the $\Gamma$-X, $\Gamma$-Z, and Z-U direction. We used the same color coding for the bands as introduced in the main text.

\bibliography{MyCollection_Suppl}